\documentclass[10pt]{amsart}
\usepackage[small]{caption}
\usepackage{amsmath,amsfonts,amscd,amssymb,epsf, epsfig}\usepackage{graphicx}

\theoremstyle{definition}

\theoremstyle{remark} 
\numberwithin{equation}{section}
\newcommand{\Z}{{\mathbb{Z}}}

\newcommand{\pa}{\partial}

\newcommand{\vep}{\varepsilon}

\begin{document}

\title[ Finite temperature Casimir effect]{ Finite temperature Casimir effect in piston geometry and its classical limit}
\author{S.C. Lim}\address{ Faculty of Engineering,
Multimedia University, Jalan Multimedia, Cyberjaya, 63100, Selangor
Darul Ehsan, Malaysia.} \email{sclim@mmu.edu.my}

\author{L.P. Teo}\address{Faculty of Information
Technology, Multimedia University, Jalan Multimedia, Cyberjaya,
63100, Selangor Darul Ehsan, Malaysia.}\email{lpteo@mmu.edu.my}

 \maketitle

\begin{abstract}
We consider the Casimir force acting on a $d$-dimensional rectangular piston due to massless scalar field with periodic, Dirichlet and Neumann boundary conditions and electromagnetic field with perfect electric conductor and perfect magnetic conductor boundary conditions. The Casimir energy in a rectangular cavity is derived  using cut-off method. It is shown that the divergent part of the Casimir energy does not contribute to the Casimir force acting on the piston, thus render an unambiguously  defined Casimir force acting on the piston. At any temperature, it is found that the Casimir force acting on the piston increases from $-\infty$ to $0$ when the  separation $a$ between the piston and the opposite wall increases from $0$ to $\infty$. This implies that the Casimir force is always an attractive force pulling the piston towards the closer wall, and the magnitude of the force gets larger as the separation $a$ gets smaller. Explicit exact expressions for the Casimir force for small and large plate separations and for low and high temperatures are computed. The limits of the Casimir force acting on the piston when some pairs of transversal plates are large are also derived.   An interesting result regarding the influence of temperature is that in contrast to the conventional result  that the leading term of the Casimir force acting on a wall of a rectangular cavity at high temperature is the Stefan--Boltzmann (or black body radiation) term which is of order $T^{d+1}$, it is found that the contributions of this term from the two regions separating the piston cancel with each other in the case of piston. The high temperature leading order term of the Casimir force acting on the piston is of order $T$, which shows that the Casimir force has a nontrivial classical $\hbar\rightarrow 0$ limit. Explicit formulas for the classical limit are computed.

\end{abstract}

\section{Introduction}
In 1948, Casimir predicted the existence of an attractive force
between two perfectly conducting parallel plates, which is due to
the zero-point fluctuation of
electromagnetic field between the plates \cite{1}. Nowadays, Casimir
effect is generally referred to similar effect due to any quantum
fields.  Since 1948, thousands of papers due to Casimir effect
have appeared in the literature, and it has gained more and more
attraction from physicists and even engineers. Many experiments have
been designed to verify the existence of the Casimir force. For
example, Mohideen et al \cite{2} used atomic force microscope to
confirm (within a few percent error) the existence of Casimir force
for plate-sphere separation between 100nm   to 900 nm. At
these length scales, Casimir force becomes non-negligible and
therefore people working in nanoscience and nanotechnology starts to
show interest in this effect. Research has been done on how to use
Casimir force to drive a nanodevice, as well as to eliminate
unpleasant effect due to Casimir force, such as adhesion or
stiction.

In the conventional calculation of Casimir energy, one sums over different modes of zero--point energies in the presence of boundaries:\begin{align}\label{eq6_11_2}E_{\text{Cas}}^0=\frac{1}{2}\sum_{\alpha}
\omega_{\alpha}.\end{align} However, this sum is divergent and regularization methods are used to obtain a finite result. One way of regularization
is by subtracting the vacuum energy due to the background infinite
space. However, this procedure does not always lead to a finite
quantity, due to the possible hypersurface divergence. As has been pointed out by  Deutsch and Candelas \cite{rr1} and Baacke and Kr\"usemann \cite{rr2}, it is impossible to compute the Casimir energy with boundary conditions to approximate the interaction of the vacuum fluctuations with the boundary material. This   has been further discussed by
Graham, Jaffe et al  in a series of papers \cite{10, 11, 12, 13}. They argued that the surface divergences cannot be removed by any renormalization of the physical parameters of the theory. In 2004,   Cavalcanti \cite{14} showed that despite the presence of the divergence in the Casimir energy, it is possible to obtain unambiguous finite Casimir force for a special geometric setup known as   piston. He showed that for a two dimensional massless
scalar field theory in a rectangular  piston, the surface
divergent terms of the Casimir force on the piston due to  the
two regions divided by the piston cancel each
other and the resulting Casimir force acting on the piston is always attractive. Since the original work of Cavalcanti, piston geometry and its variants have attracted considerable interest.
In \cite{15, 16}, it was shown that for electromagnetic field with
perfect electric conductor conditions inside three dimensional rectangular
 piston, the surface divergent terms are also canceled and
the resulting Casimir force acting on the piston is  attractive. The Casimir piston for three
dimensional electromagnetic field with prefect conductor conditions
are studied further in \cite{17, 18, 19}, where pistons with
arbitrary cross sections are considered. In \cite{25}, the
rectangular   piston for massless scalar field with Dirichlet boundary conditions in three
dimensions was discussed in detail. In \cite{26}, it was proved that
for massless scalar field with Dirichlet and Neumann boundary conditions, the hypersurface divergent terms of  the Casimir force due to the
two regions separating a $d$-dimensional rectangular
  piston  always cancel each other, and the Casimir force
is always attractive. The Casimir force on rectangular piston due to
electromagnetic field with perfect magnetic conductor conditions are
computed and discussed by Edery and Marachevsky \cite{29}. In another recent
work \cite{30}, the Casimir force for   massless scalar field
with Dirichlet boundary conditions on a rectangular piston in the
spacetime with extra compactified dimensions  was discussed. In
\cite{20}, it was shown that the Casimir force between two
(nonmagnetic) dielectric bodies which are related by reflection is
always attractive. This result was generalized by Bachas \cite{21}  who showed that reflection positivity implies that the force
between any mirror pair of charge-conjugate probes of the quantum
vacuum is attractive. The attractive nature of the Casimir force
will create undesirable effects such as the collapse of a nano
device -- an effect known as stiction \cite{22, 23}. Therefore, it
becomes desirable to search for circumstances where the Casimir
force can be made less attractive, or even repulsive. In \cite{24},
Barton showed that for a thin piston with weakly reflecting
dielectrics, the Casimir force at small separations is attractive,
but turn to repulsive  as the separation increases. In \cite{27}, it
was shown that in one, two or three dimensions, if one surface
assumes Dirichlet boundary condition and the other assumes Neumann
boundary condition, then the Casimir force on a rectangular piston
is repulsive. Another scenario that leads to repulsive Casimir force
was discussed in \cite{28}. We would like to remark that a
scenario similar to piston geometry has been considered by Reuter and Dittrich in 1985 \cite{32} in the context  as a mechanism to regularize the
Casimir energy.   Recently, a similar mechanism is
used to compute the Dirichlet Casimir effect for $\phi^4$ theory in
(3+1) dimensions \cite{33}.

 In this paper, we  study the Casimir effect of massless scalar field and electromagnetic field in a $d$-dimensional rectangular piston at   finite temperature. Different boundary conditions are considered including periodic, Dirichlet and Neumann boundary conditions for massless scalar field and
 perfect electric conductor (PEC)   and
perfect magnetic conductor (PMC) boundary conditions for electromagnetic field.  In this paper, we propose an alternative
method to calculate the  cut-off dependent   Casimir energy. We notice that the divergent part of the Casimir energy is linear in each of the variables $L_1, \ldots, L_d$ --- the lengths  of the rectangular cavity. This immediately implies that the divergent terms do not contribute to the Casimir force acting on the piston. Therefore  the Casimir force acting on the piston  is finite and unambiguous without any renormalization.   The thermal effect of the Casimir force   has so far
only  been  considered for electromagnetic field with PEC boundary conditions in three
dimensions \cite{16, 18, 19}. However, \cite{16} does not give a correct high temperature behavior of the Casimir force.  In this paper, we derive for $d$-dimensional piston geometry explicit and exact formulas of the Casimir force which can be used to study the high temperature and low temperature behaviors of the Casimir force. We show that in the high temperature regime, the black body radiation contributions to the Casimir force from the two regions separating the piston cancel with each other. The leading order term of the Casimir force is of order $T$, which shows that the Casimir force has a classical limit.

In this paper,  we choose the units where $\hbar=c=k=1$.
%We are going to give a rigorous proof that the Casimir force is
%always negative (attractive) and its magnitude decreases as the
%separation between the piston and the opposite wall increases.

\section{Casimir energy by exponential cut-off method}\label{sec2}

Consider a
$d$-dimensional rectangular piston, which is a
rectangular cavity  $[0, L_1]\times [0,L_2]\times \ldots [0,L_d]$
separated by a hyperplane $x_1=a$ (the piston) to two regions: the
  region I of dimension $[0, a]\times [0,L_2]\times \ldots [0,L_d]$ and the
  region II of dimension  $[a, L_1]\times [0,L_2]\times \ldots [0,L_d]$. We are mainly concerned with the case where $L_1\rightarrow\infty$ which implies that
region II is opened. The case where $L_1$ remains finite will be discussed in a later section.  The two and three dimensional
rectangular pistons are illustrated in Figure 1.  Casimir force acting on the piston due to massless scalar field with
periodic (P), Dirichlet (D) and Neumann (N) boundary conditions
(b.c.), as well as electromagnetic field with perfect electric conductor (PEC)
b.c.~ and perfect magnetic conductor  (PMC) b.c.~ will be computed.
\begin{figure}\centering \epsfxsize=.49\linewidth
\epsffile{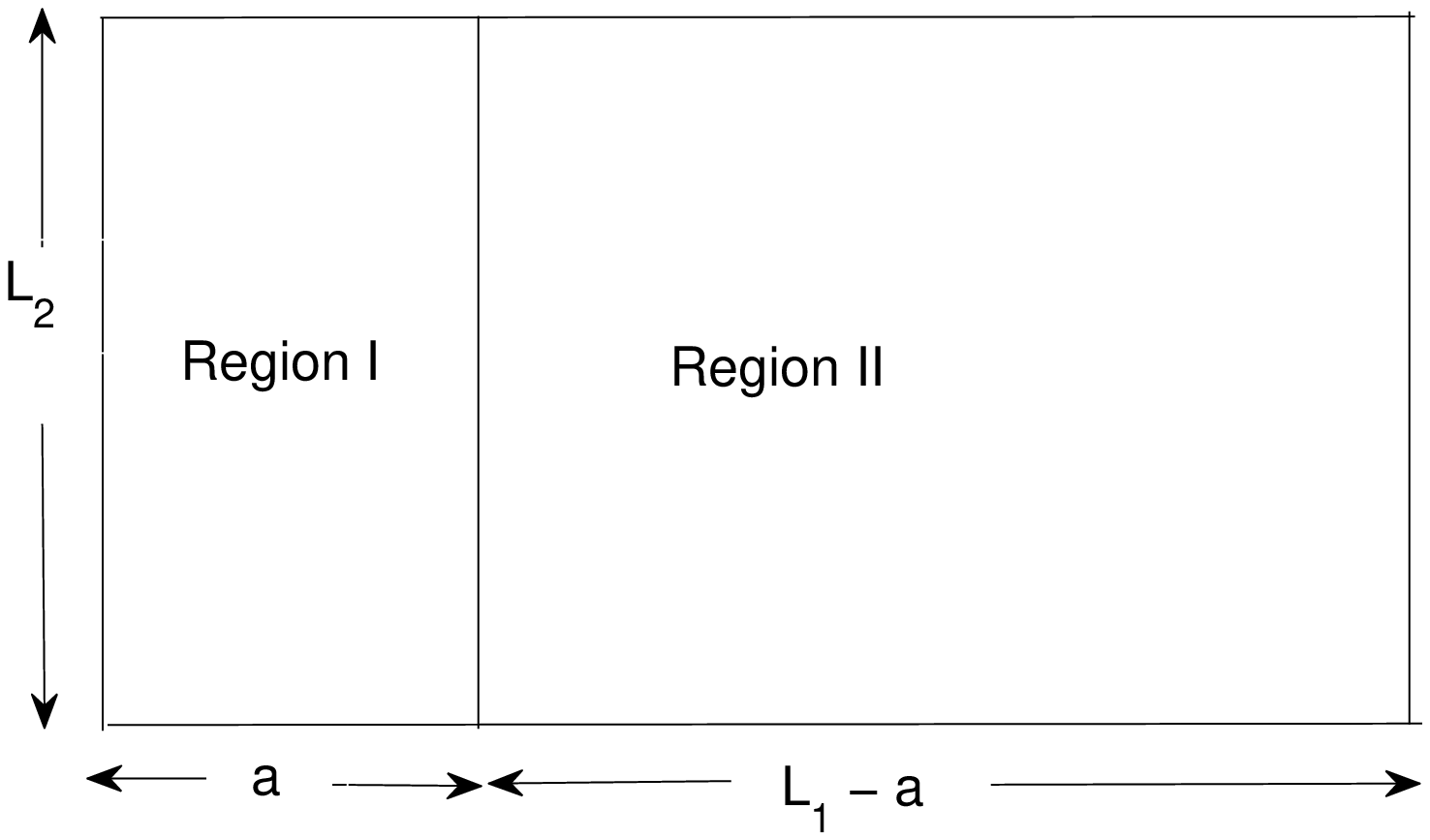} \epsfxsize=.49\linewidth
\epsffile{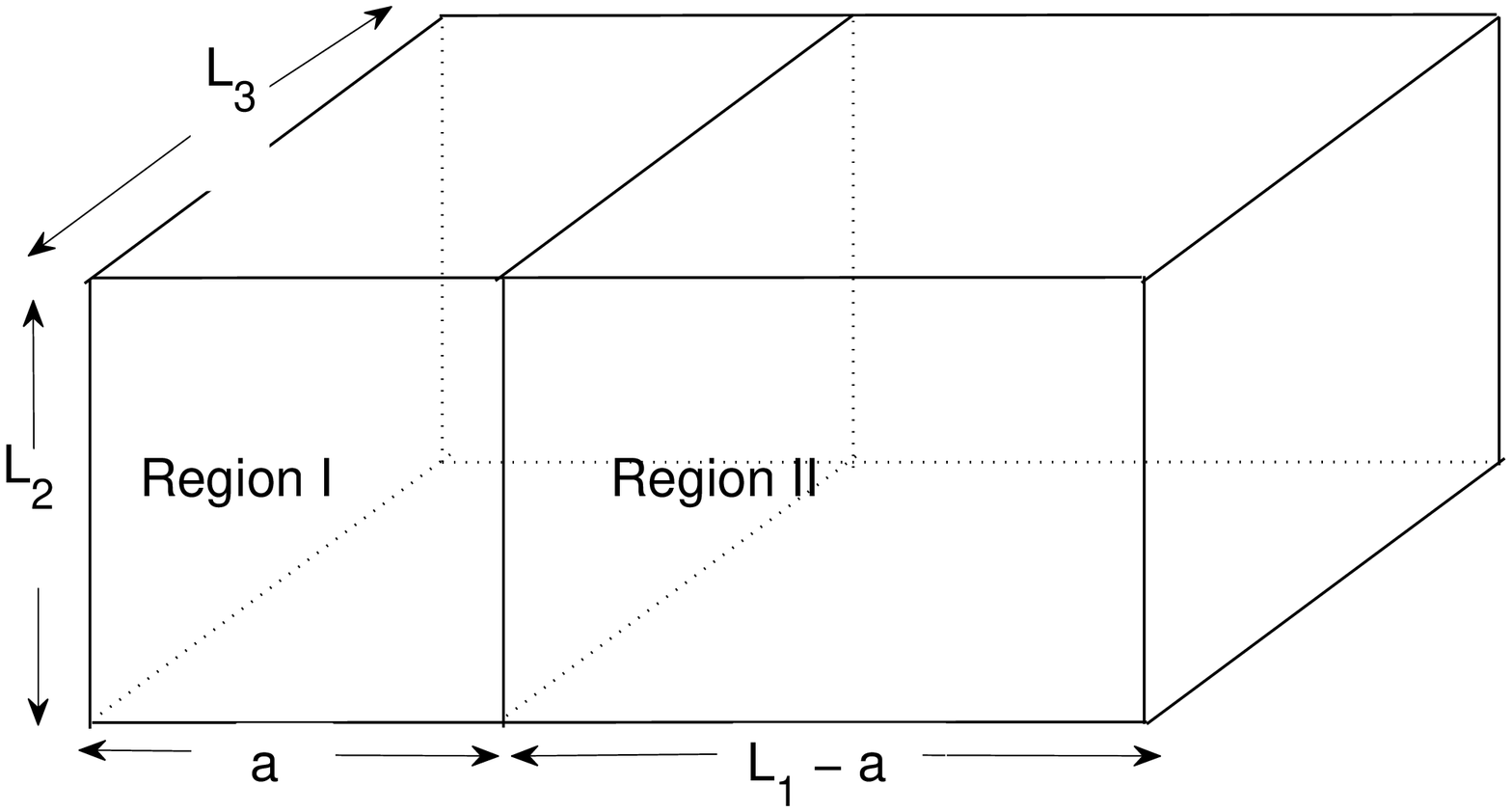}\caption{The two and three dimensional
rectangular pistons}\end{figure}

We first recall some basic definitions. The finite temperature Casimir energy is naively the sum of
the zero temperature Casimir energy \eqref{eq6_11_2} and the thermal correction:
\begin{align*}
E_{\text{Cas}}= E_{\text{Cas}}^0+\Delta E_{\text{Cas}}= E_{\text{Cas}}^0+T\sum_{\omega_\alpha\neq 0}
\log\left(1-e^{-\omega_{\alpha}/T}\right)=-T \log
\prod_{\omega_{\alpha}\neq
0}\frac{e^{-\omega_{\alpha}/(2T)}}{1-e^{-\omega_{\alpha}/T}}.
\end{align*}  For the piston system, the Casimir energy is the sum of the Casimir energies of regions I and II and the Casimir energy of the region outside the rectangular cavity:
\begin{align}\label{eq10_8_2}
E_{\text{Cas}}=E_{\text{Cas}}^{\text{I}}+E_{\text{Cas}}^{\text{II}}+E_{\text{Cas}}^{\text{out}}.
\end{align}Although each of these terms is divergent, we are going to see below that the Casimir force acting on the piston   defined by
\begin{align}\label{eq6_12_12}
F_{\text{Cas}}= -\frac{\pa}{\pa
a} E_{\text{Cas}}= -\frac{\pa}{\pa
a} \left(E_{\text{Cas}}^{\text{I}}+E_{\text{Cas}}^{\text{II}}\right)
\end{align}is finite without any renormalization. The term $E_{\text{Cas}}^{\text{out}}$ has drop  out from the last equality of \eqref{eq6_12_12} since it is independent of $a$.

 Notice that
$E_{\text{Cas}}^{\text{I}}=E_{\text{Cas}}(a, L_2, \ldots, L_d)$
and $E_{\text{Cas}}^{\text{II}}=E_{\text{Cas}}(L_1-a, L_2, \ldots,
L_d)$, where $E_{\text{Cas}}(L_1, L_2, \ldots, L_d)$ is the Casimir
energy of the rectangular cavity $[0, L_1]\times [0,L_2]\times
\ldots [0,L_d]$. As was proved in \cite{39, 34}, the Casimir energy
$E_{\text{Cas}}(L_1, L_2, \ldots, L_d)$ for massless scalar field
with Pb.c., Db.c.~ and Nb.c.~ and for electromagnetic
field with PEC b.c.~~ and PMC b.c.~~ are related to each other by the following
\emph{linear} relations:
\begin{align}\label{eq6_12_1}
E_{\text{Cas}}^{D/N}(L_1, \ldots, L_d)=&2^{-d}\sum_{j=1}^{d}(\mp
1)^{d-j} \sum_{1\leq m_1<\ldots<m_j\leq
d}E_{\text{Cas}}^{P}(2L_{m_1}, \ldots, 2L_{m_j}),
\end{align}
\begin{align}\label{eq6_12_2}
E_{\text{Cas}}^{\text{PEC}}(L_1, \ldots,
L_d)=(d-1)E_{\text{Cas}}^D(L_1,\ldots,L_d)+\sum_{j=1}^d
E_{\text{Cas}}^{D}(L_1,\ldots, L_{j-1}, L_{j+1}, \ldots, L_d),
\end{align}and
\begin{align}\label{eq6_12_3}
E_{\text{Cas}}^{\text{PMC}}(L_1, \ldots,
L_d)=&\sum_{j=2}^{d}(j-1)\sum_{1\leq m_1<\ldots <m_j\leq
d}E_{\text{Cas}}^{D}\left(s; L_{m_1},\ldots,
L_{m_j}\right).
\end{align}
Notice that when the dimension $d$ is equal to one, there are no electromagnetic field satisfying either PEC b.c.~ or PMC b.c.; whereas the Casimir energy for massless scalar field with Db.c.~ and Nb.c.~ is related to the Casimir energy for massless scalar field with Pb.c.~ by the simple relation $E_{\text{Cas}}^{D/N}(a) = (1/2) E_{\text{Cas}}^{P}(2a)$. In the following discussions, when we mention massless scalar field with Db.c.~ and Nb.c.~ or electromagnetic field with PEC b.c.~ and PMC b.c., we implicitly assume that the space dimension $d$ is larger than two.

For a massless scalar field with Pb.c.~ in a rectangular
cavity $[0, L_1]\times [0,L_2]\times \ldots [0,L_d]$, the eigenfrequencies are
\begin{align*}
 \sqrt{\sum_{j=1}^d \left(\frac{2\pi
k_j}{L_j}\right)^2}, \hspace{1cm}\mathbf{k}=(k_1,\ldots, k_d)\in
\Z^d.
\end{align*} Let
\begin{align}\label{eq6_12_4}
E_{\text{Cas}}^{P,0} (\lambda; L_1, \ldots,
L_d)=&\frac{1}{2}\sum_{\mathbf{k}\in \Z^d\setminus\{0\}}\sqrt{\sum_{j=1}^d
\left(\frac{2\pi
k_j}{L_j}\right)^2}\exp\left(-\lambda\sqrt{\sum_{j=1}^d
\left(\frac{2\pi
k_j}{L_j}\right)^2}\right)\\=&-\frac{1}{2}\frac{\pa}{\pa
\lambda}K(\lambda; L_1, \ldots, L_d)\nonumber
\end{align}be the $\lambda$-dependent zero temperature Casimir
energy. Here the function $K(\lambda; L_1, \ldots, L_d)$ is defined
by
\begin{align*}
K(\lambda; L_1, \ldots, L_d)=\sum_{\mathbf{k}\in
\Z^d\setminus\{0\}}\exp\left(-\lambda\sqrt{\sum_{j=1}^d \left(\frac{2\pi
k_j}{L_j}\right)^2}\right).
\end{align*}
  Using the  formula
\begin{align*}
e^{-z} =\frac{1}{2\pi i} \int_{c-i\infty}^{c+i\infty} \Gamma(w)
z^{-w}dw, \hspace{1cm} \text{Re}\;c>0,
\end{align*}we find that
\begin{align}\label{eq6_12_5}
K(\lambda; L_1, \ldots, L_d)=&\frac{1}{2\pi
i}\int_{c-i\infty}^{c+i\infty} \sum_{\mathbf{k}\in \Z^d\setminus\{0\}}\Gamma(w)
\lambda^{-w}\left(\sum_{j=1}^d \left[\frac{2\pi
k_j}{L_j}\right]^2\right)^{-\frac{w}{2}}dw\\
=&\frac{1}{2\pi i}\int_{c-i\infty}^{c+i\infty} \Gamma(w)\lambda^{-w}
Z_{d}\left(\frac{w}{2};
\frac{2\pi}{L_1},\ldots,\frac{2\pi}{L_d}\right),\hspace{1cm}\text{Re}\;
c > \frac{d}{2},\nonumber
\end{align}where
$Z_{d}(s; a_1,\ldots, a_d)$ is the homogeneous Epstein zeta function
\cite{35, 36} defined by the infinite series
\begin{align}\label{eq10_8_3}
Z_{d}(s; a_1,\ldots, a_d)=\sum_{\mathbf{k}\in\Z^d\setminus\{0\}}
\left( \sum_{j=1}^{d}[a_jk_j]^2\right)^{-s}
\end{align}when $\text{Re}\; s>d/2$.
  The function $Z_{d}\left(\frac{w}{2};
\frac{2\pi}{L_1},\ldots,\frac{2\pi}{L_d}\right)$ has a meromorphic continuation to $\mathbb{C}$ with a simple
pole at $w=d$ with residue
$$\frac{1}{2^{d-1}\pi^{d/2}\Gamma\left(\frac{d}{2}\right)}\prod_{j=1}^d
L_j.$$By shifting the contour of integration in \eqref{eq6_12_5}
from the line $\text{Re}\; w= c, c>d/2$ to a line $\text{Re} \; w=
-2-\vep, \vep>0$, we  find that
\begin{align}\label{eq6_12_9}
E_{\text{Cas}}^{P,0} (\lambda; L_1, \ldots,
L_d)=\frac{\Gamma(d+1)}{2^{d}\pi^{d/2}\Gamma\left(\frac{d}{2}\right)}\left[\prod_{j=1}^d
L_j\right]\lambda^{-d-1}+\frac{1}{2}Z_{d}\left(-\frac{1}{2};
\frac{2\pi}{L_1}, \ldots,\frac{2\pi}{L_d}\right) +O(\lambda).
\end{align}Using
 the functional equation (see e.g. \cite{ 35, 36, 4, 5, 6})
\begin{align}\label{eq6_12_7}
\pi^{-s}\Gamma(s)Z_{d}(s; a_1, \ldots , a_d)
=\frac{\pi^{s-\frac{d}{2}}}{\left[\prod_{j=1}^da_j\right]}\Gamma\left(\frac{d}{2}-s\right)Z_{d}\left(\frac{d}{2}-s;
\frac{1}{a_1},\ldots, \frac{1}{a_d}\right),
\end{align}we find that  the   cut-off dependent finite
temperature Casimir energy is given by
\begin{align}\label{eq6_12_13} E_{\text{Cas}}^{P} (\lambda;
L_1, \ldots,
L_d)=\frac{\Gamma(d+1)}{2^{d}\pi^{d/2}\Gamma\left(\frac{d}{2}\right)}\left[\prod_{j=1}^d
L_j\right]\lambda^{-d-1}+E_{\text{Cas, reg}}^{P} ( L_1, \ldots,
L_d)+O(\lambda),
\end{align}where the regularized finite temperature Casimir energy
$E_{\text{Cas, reg}}^{P} ( L_1, \ldots, L_d)$
is given by (see \cite{39}):
\begin{equation}\label{eq10_8_1}\begin{split}
E_{\text{Cas, reg}}^{P} ( L_1, \ldots,
L_d)=& -\frac{\Gamma\left(\frac{d+1}{2}\right)}{2\pi^{\frac{d+1}{2}}}\left[\prod_{j=1}^d
L_j\right] Z_d\left(\frac{d+1}{2}; L_1, \ldots, L_d\right)\\&+T  \sum_{\mathbf{k}\in\Z^{d}\setminus\{0\}}
\log\left(1-\exp\left(-\frac{2\pi }{T}\sqrt{\sum_{j=1}^d \left(\frac{k_j}{L_j}\right)^2}\right)\right)\\
=& -\frac{T}{2}
Z_{d+1}'\left(0; T, \frac{1}{L_1}, \ldots,
\frac{1}{L_d}\right)-T\log\frac{2\pi}{T}.
\end{split}\end{equation}
The cut-off
dependent finite temperature Casimir energy for massless scalar field with Db.c.~ and Nb.c.~ and electromagnetic field with PEC b.c.~ and PMC b.c.~
can be computed using formulas \eqref{eq6_12_1}, \eqref{eq6_12_2} and
\eqref{eq6_12_3}. It is easy to deduce from \eqref{eq6_12_13},
\eqref{eq6_12_1}, \eqref{eq6_12_2} and \eqref{eq6_12_3} that in all
cases, the cut-off dependent Casimir energy can be written as a sum
of the $\lambda\rightarrow 0^+$ divergent term and the regularized
term:
\begin{align*}
E_{\text{Cas}}(\lambda; L_1, \ldots, L_d) =E_{\text{Cas,
div}}(\lambda; L_1, \ldots, L_d) +E_{\text{Cas, reg}}( L_1, \ldots,
L_d)+O(\lambda).
\end{align*}Although the divergent term for the massless scalar field with
Pb.c.~ only depends on the bulk volume, we can see from
\eqref{eq6_12_1}, \eqref{eq6_12_2} and \eqref{eq6_12_3} that for massless scalar field with Db.c.~ and Nb.c.~ and electromagnetic field with PEC b.c.~ and PMC b.c., the divergent terms depend on the area of lower
dimensional hypersurfaces. As mentioned in the introduction, it has been argued by several authors \cite{rr1, rr2, 10, 11, 12, 13} that these hypersurface divergences cannot be removed by renormalization of physical parameters.
 However, we notice that regarded as a
function of $L_1$,  \emph{the divergent term is linear in $L_1$},
i.e.
\begin{align}\label{eq6_12_14}
E_{\text{Cas, Div}}(\lambda; L_1, \ldots, L_d) = e_1(\lambda, L_2,
\ldots, L_d) L_1+e_0(\lambda; L_2, \ldots, L_d).
\end{align}

Here we want to remark that the above calculations is an idealization of the more physical description of the interaction between the quantum field and an external potential, which goes to zero away from the boundary surfaces. For a more rigorous treatment, one can consider the approach of \cite{10}
where the imposed boundary conditions are approximated by adding a $\delta$-type background potential to the quantum field  which is concentrated on the boundaries. In our idealization where the piston is assumed to be a perfect rigid rectangular cavity partitioned into two regions by a perfect rigid piston with negligible thickness, the Casimir force acting on the piston would be independent of the approach used to compute the Casimir energy, as is already pointed out by \cite{14}. However, we would like to emphasize that we do not claim to have renormalized the Casmir energy in a physical way. Instead, our point is that for an idealized piston, the divergence terms which plague the calculations of Casimir energy can be ignored since they would not contribute to the Casimir force acting on the piston (see next section). But this only holds in the idealized situation. For more details about physical situations that would invalidate our assumptions, one can refer to \cite{10} and a more recent preprint \cite{nn1}.
\section{Casimir force on $d$-dimensional rectangular pistons}\label{sec3}
Since the $\lambda\rightarrow 0^+$ divergent term of the Casimir
energy for a $d$-dimensional rectangular cavity is linear in $L_1$
\eqref{eq6_12_14}, it follows that for the Casimir energy of the
piston system \eqref{eq10_8_2}, the $\lambda\rightarrow 0^+$ divergent term depends  only on $L_1$,
and not on $a$. Therefore the $\lambda\rightarrow 0^+$ divergent
part of the Casimir energy do not contribute to the Casimir force acting on the piston. Consequently, the Casimir force can be computed by using only the regularized Casimir energy:
\begin{align}\label{eq7_11_1}
F_{\text{Cas}}(a; L_2, \ldots, L_d)=-\lim_{L_1\rightarrow
\infty}\frac{\pa}{\pa a}\Bigl\{E_{\text{Cas, reg}}(a, \ldots,
L_d)+E_{\text{Cas, reg}}(L_1-a, \ldots, L_d)\Bigr\}.
\end{align}

 For massless scalar field with Pb.c., we obtain from eq. \eqref{eq7_22_3} that the regularized Casimir force acting on the piston due to the vacuum fluctuation of the field in region II is given by
\begin{align}\label{eq7_15_10}
F_{\text{Cas}}^{P, \text{II}}(a; L_2, \ldots, L_d) =& \pi T
 Z_d\left(-\frac{1}{2}; \frac{1}{L_2}, \ldots, \frac{1}{L_d},
T\right)\\=&-\frac{\Gamma\left(\frac{d+1}{2}\right)}{2\pi^{\frac{d+1}{2}}} \left[\prod_{j=2}^dL_j\right]Z_d\left(\frac{d+1}{2}; L_2, \ldots, L_d,
\frac{1}{T}\right);\nonumber
\end{align}
and the total Casimir force acting on the piston is
\begin{align}\label{eq7_14_1}
F_{\text{Cas}}^P(a; L_2, \ldots, L_d) =&-\frac{T}{a}-2\pi
T\sum_{k_1=1}^{\infty} \sum_{(k_2, \ldots,
k_{d}, l)\in\Z^{d}\setminus\{0\}}\left(\sum_{j=2}^d
\left(\frac{k_j}{L_j}\right)^2 +(lT)^2\right)^{\frac{1}{2}}\\
&\times \exp\left(-2\pi k_1 a \sqrt{\sum_{j=2}^d
\left(\frac{k_j}{L_j}\right)^2 +(lT)^2}\right)\nonumber
\end{align}for massless scalar field with Pb.c. For massless scalar field with Db.c.~ and Nb.c., \eqref{eq6_12_12} and \eqref{eq6_12_1} give
\begin{align}\label{eq7_14_9}
F_{\text{Cas}}^{D/N}(a; L_2, \ldots, L_d)=2^{-d+1}\sum_{j=1}^{d} (\mp)^{d-j}\sum_{2\leq m_1<\ldots <m_{j-1}\leq d}
F_{\text{Cas}}^P(2a; 2L_{m_1}, \ldots, 2L_{m_{j-1}}).
\end{align}For $j=1$, the second summation is a single term $F_{\text{Cas}}^P(2a)$. Using \eqref{eq7_14_1}, we find that the Casimir force on the piston for massless scalar field with Db.c.~ and Nb.c.~ can be written respectively as
\begin{align}\label{eq7_14_2}
&F_{\text{Cas}}^D(a; L_2, \ldots, L_d) =-\pi
T \sum_{k_1=1}^{\infty} \sum_{(k_2, \ldots,
k_{d})\in\mathbb{N}^{d-1}}\sum_{l=-\infty}^{\infty} \left(\sum_{j=2}^d
\left(\frac{k_j}{L_j}\right)^2 +(2lT)^2\right)^{\frac{1}{2}}\\
&\hspace{4cm}\times \exp\left(-2\pi k_1 a \sqrt{\sum_{j=2}^d
\left(\frac{k_j}{L_j}\right)^2 +(2lT)^2}\right)\nonumber
\end{align}and
\begin{align}\label{eq7_14_3}
&F_{\text{Cas}}^N(a; L_2, \ldots, L_d) =-\frac{T}{2a}-\pi
T\sum_{k_1=1}^{\infty} \sum_{(k_2, \ldots, k_d, l) \in  \left[(\mathbb{N}\cup\{0\}
)^{d-1}\times \Z \right]\setminus \{0\}}
\\
&\times \left(\sum_{j=2}^d
\left(\frac{k_j}{L_j}\right)^2 +(2lT)^2\right)^{\frac{1}{2}}\exp\left(-2\pi k_1 a \sqrt{\sum_{j=2}^d
\left(\frac{k_j}{L_j}\right)^2 +(2lT)^2}\right).\nonumber
\end{align}For electromagnetic field under PEC b.c.~ and PMC b.c., \eqref{eq6_12_2} and \eqref{eq6_12_3} give
\begin{align}\label{eq7_14_4}
&F_{\text{Cas}}^{PEC} (a; L_1, \ldots, L_d)= (d-1)F_{\text{Cas}}^D(a; L_2, \ldots, L_d) +\sum_{j=2}^{d} F_{\text{Cas}}^D(a; L_2, \ldots, L_{j-1}, L_{j+1},
\ldots, L_{d});
\end{align}and\begin{align}\label{eq7_14_5}
F_{\text{Cas}}^{PMC} (a; L_1, \ldots, L_d)=& \sum_{j=2}^{d} (j-1)\sum_{2\leq m_1<\ldots<m_{j-1}\leq d}F_{\text{Cas}}^D(a; L_{m_1}, \ldots, L_{m_{j-1}}).
\end{align}  Using  these formulas and the formula \eqref{eq7_14_2}, we conclude that
\begin{align}\label{eq7_15_1}
&F_{\text{Cas}}^{PEC/PMC}(a; L_2, \ldots, L_d) =-\pi
T\sum_{k_1=1}^{\infty} \sum_{(k_2, \ldots,
k_{d})\in(\mathbb{N}\cup\{0\})^{d-1}}\Lambda^{PEC/PMC}(k_2,\ldots, k_d)\\&\times\sum_{l=-\infty}^{\infty}\left(\sum_{j=2}^d
\left(\frac{k_j}{L_j}\right)^2 +(2lT)^2\right)^{\frac{1}{2}} \exp\left(-2\pi k_1 a \sqrt{\sum_{j=2}^d
\left(\frac{k_j}{L_j}\right)^2 +(2lT)^2}\right)-\delta^{PEC/PMC}\frac{T}{2a};\nonumber
\end{align}
where
\begin{align}\label{eq7_15_11}
\Lambda^{PEC}(k_2,\ldots, k_d)=\begin{cases} d-1, \hspace{0.5cm}&\text{if all $k_i$, $i=2, \ldots, d$ are nonzero},\\
1, &\text{if exactly one of the $k_i$ is zero},\\
0, &\text{if more than one of the $k_i$ is zero};\end{cases}
\end{align}
\begin{align}\label{eq7_15_12}
\Lambda^{PMC}(k_2,\ldots, k_d)=  j, \hspace{0.5cm}&\text{if   exactly $j$ of the $k_i$, $i=2, \ldots, d$ is nonzero};
\end{align}and $\delta^{PEC} =1$ if and only if $d=2$ and $\delta^{PMC}\equiv 0$.

Notice that since the Bessel function $K_{\nu}(z)$ is positive for any positive $z$, we immediately obtain from \eqref{eq7_14_1},
\eqref{eq7_14_2},  \eqref{eq7_14_3} and \eqref{eq7_15_1} that \emph{for massless scalar field with periodic, Dirichlet and Neumann boundary conditions and for electromagnetic field with PEC and PMC boundary conditions, the Casimir force acting on the piston always has negative sign, and therefore is attractive. This holds for any dimension and size of the piston as well as for any temperature. }

By taking derivative of the Casimir force with respect to $a$, one finds from
\eqref{eq7_14_1}, \eqref{eq7_14_2}, \eqref{eq7_14_3} and \eqref{eq7_15_1}  that the derivative of the Casimir force with respect to $a$ is always positive. Therefore, we can conclude that \emph{for either massless scalar field with periodic, Dirichlet or Neumann boundary conditions or electromagnetic field with PEC and PMC boundary conditions, the Casimir force is always an increasing function of $a$}. Since the Casimir force is always negative, this implies that \emph{the magnitude of the Casimir force is always decreased when $a$ is increased.} To the best of our knowledge, this
is the first time such results are obtained analytically.

\section{Asymptotic behaviors of the Casimir force for large and small plate separations}

The formulas \eqref{eq7_14_1}, \eqref{eq7_14_2}, \eqref{eq7_14_3} and \eqref{eq7_15_1}  are ideal for studying the large--$a$ and small $L_j$, $2\leq j\leq d$ behaviors of the Casimir force. In particular,  the leading order terms of the Casimir force when $a\rightarrow \infty$ are given respectively by
\begin{align*}
F_{\text{Cas}}^P(a; L_2, \ldots, L_d) \sim & -\frac{T}{a}-4\pi T\min\{L_2^{-1},\ldots, L_d^{-1}, T\} e^{-2\pi a\min\{L_2^{-1},
\ldots, L_d^{-1}, T\}},
\end{align*}\begin{align}\label{eq7_15_3}F_{\text{Cas}}^D(a; L_2, \ldots, L_d) \sim& - \pi T\sqrt{\sum_{j=2}^dL_j^{-2}} e^{-2\pi a \sqrt{\sum_{j=2}^d L_j^{-2}}},\end{align}\begin{align*}
F_{\text{Cas}}^N(a; L_2, \ldots, L_d) \sim &-\frac{T}{2a}-\pi T\min\{L_2^{-1},\ldots, L_d^{-1}, 2T\} e^{-2\pi a\min\{L_2^{-1},
\ldots, L_d^{-1}, 2T\}},\end{align*}\begin{align}\label{eq7_15_4}
F_{\text{Cas}}^{PEC}(a; L_2, \ldots, L_d) \sim& -\delta^{PEC}\frac{T}{2a}-\pi T\min\left\{\sqrt{\sum_{\substack{l=2\\l\neq j}}^dL_l^{-2}}\right\}_{j=2}^d \exp\left(-2\pi a \min\left\{\sqrt{\sum_{\substack{l=2\\l\neq j}}^dL_l^{-2}}\right\}_{j=2}^d\right)\end{align}\begin{align}\label{eq7_15_5}
F_{\text{Cas}}^{PMC}(a; L_2, \ldots, L_d) \sim &-\pi T\min\{L_2^{-1},\ldots, L_d^{-1}\} e^{-2\pi a\min\{L_2^{-1},
\ldots, L_d^{-1}\}}.
\end{align}The case where $T=0$ will be considered in the next section. These asymptotic behaviors imply that when $a\rightarrow \infty$, the magnitude of the Casimir force tends to zero. Moreover, for massless scalar field with Pb.c.~ and Nb.c.~ and electromagnetic field with PEC b.c.~ in $d=2$ dimension, the Casimir force tends to zero polynomially, in the order $a^{-1}$. However, for massless scalar field with Db.c.~ and for electromagnetic field with PEC b.c.~ and PMC b.c., the magnitude of the Casimir force decays to zero exponentially fast.

For the Casimir force when the   plate separation $a$ is small,    \eqref{eq7_14_8}   and \eqref{eq7_15_10} give us
\begin{align}\label{eq7_14_10}
&F_{\text{Cas}}^P(a; L_2, \ldots, L_d) =
-\frac{d\Gamma\left(\frac{d+1}{2}\right)}{\pi{^\frac{d+1}{2}} a^{d+1}} \left[\prod_{j=2}^d L_i\right]\zeta_R(d+1)+\pi TZ_d\left( -\frac{1}{2}; \frac{1}{L_2}, \ldots, \frac{1}{L_d}, T\right)\\&+\frac{4\pi}{ a^{\frac{d+4}{2}}}\left[\prod_{j=2}^d L_i\right]\sum_{k_1=1}^{\infty} \sum_{(k_2, \ldots, k_d, l)\in \Z^{d}\setminus\{0\}}
k_1^{\frac{d+2}{2}}\left(\sum_{j=2}^d (k_jL_j)^2 + \left(\frac{l}{T}\right)^2\right)^{-\frac{d-2}{4}}\nonumber\\&\times K_{\frac{d-2}{2}}\left( \frac{2\pi k_1}{a}
\sqrt{\sum_{j=2}^d (k_jL_j)^2 + \left(\frac{l}{T}\right)^2}\right).\nonumber
\end{align}When $a\rightarrow 0^+$, the first term tends to infinity, of order $a^{-d-1}$. The second term is $O(a^0)$ and the third term tends to 0
exponentially fast. In other words, when the plate separation $a$ is small, the leading term of the Casimir force for massless scalar field with Pb.c.~ is
\begin{align}\label{eq7_15_6}
F_{\text{Cas}}^P(a; L_2,\ldots, L_d) \sim -\frac{d\Gamma\left(\frac{d+1}{2}\right)}{\pi{^\frac{d+1}{2}} a^{d+1}} \left[\prod_{j=2}^d L_i\right]\zeta_R(d+1)+O(a^0).
\end{align}For massless scalar field with Db.c.~ and Nb.c.~ and electromagnetic field with PEC b.c.~ and PMC b.c., the corresponding expression for the Casimir force when the  separation $a$ is small can be found by substituting \eqref{eq7_14_10} into \eqref{eq7_14_9}, \eqref{eq7_14_4} and \eqref{eq7_14_5}.  In particular, we find  that when $a$ is small, the asymptotic behaviors of the Casimir force are given respectively by
\begin{align*}
F_{\text{Cas}}^{D/N}(a; L_2, \ldots, L_d)\sim -\frac{1}{2^{d+1}}\sum_{j=1}^d(\mp)^{d-j} \frac{j\Gamma\left(\frac{j+1}{2}\right)\zeta_R(j+1)}{ \pi^{\frac{j+1}{2}}a^{j+1}}
S_j+O(a^0),
\end{align*}and\begin{align*}
F_{\text{Cas}}^{PEC/PMC}(a; L_2, \ldots, L_d)\sim -\frac{1}{2^{d+1}}\sum_{j=1}^d(\mp)^{d-j} (2j-d-1) \frac{j\Gamma\left(\frac{j+1}{2}\right)\zeta_R(j+1)}{\pi^{\frac{j+1}{2}}a^{j+1}}
S_j+O(a^0);
\end{align*}where
\begin{align}\label{eq7_25_1}S_j =\sum_{2\leq m_1 <\ldots<m_{j-1}\leq d} L_{m_1}\ldots L_{m_{j-1}}.\end{align} Therefore for massless scalar field with Db.c.~ or Nb.c., the  small-$a$ leading term of the Casimir force is $$- \frac{d\Gamma\left(\frac{d+1}{2}\right)\zeta_R(d+1)}{2^{d+1}\pi{^\frac{d+1}{2}} a^{d+1}} \left[\prod_{j=2}^d L_i\right],$$ which is $2^{d+1}$ times smaller than the leading term for the massless scalar field with Pb.c. For electromagnetic field with PEC b.c.~ and PMC b.c., the   small-$a$  leading term  is $$-(d-1)\frac{d\Gamma\left(\frac{d+1}{2}\right)\zeta_R(d+1)}{2^{d+1}\pi{^\frac{d+1}{2}} a^{d+1}} \left[\prod_{j=2}^d L_i\right],$$ which is $\frac{2^{d+1}}{d-1}$ times smaller than the leading term for the massless scalar field with Pb.c.~ and $d-1$ times larger than the leading term for the massless scalar field with Db.c.

Notice that these leading order terms are independent of the temperature $T$.
Therefore we conclude that \emph{at low temperature,  the effect of temperature to Casimir force is insignificant when the plate separation $a$ is small.}  On the other hand, the  leading order terms also show that the Casimir force tends to $-\infty$ when $a\rightarrow 0^+$. Combining with the fact that the Casimir force tends to $0$ as $a\rightarrow \infty$ and the Casimir force is an increasing function of $a$, we conclude that \emph{for massless scalar field with Pb.c., Db.c.~ and Nb.c.~ and electromagnetic field with PEC b.c.~ and PMC b.c., the Casimir force always increases from $-\infty$ to $0$ as the separation $a$ increases from 0 to $\infty$}.

\section{Low temperature and High temperature expansions of the Casimir force}

As discussed in \cite{39}, the low temperature expansion of the regularized Casimir energy inside a rectangular cavity  is just the zero temperature Casimir energy plus the temperature correction. In the case of massless scalar field with Pb.c., it is given explicitly by \eqref{eq10_8_1}.
Together with \eqref{eq7_15_10} and   \eqref{eq7_15_18}, we find that the low temperature expansion of the Casimir force acting  on the piston is given by
\begin{align}\label{eq7_15_16}
&F_{\text{Cas}}^{P} (a; L_2, \ldots, L_d) = F_{\text{Cas}}^{P,  T=0} (a; L_2, \ldots, L_d)+\frac{4\pi}{a^3}
\sum_{l=1}^{\infty}\sum_{k_1=1}^{\infty}\sum_{(k_2,\ldots, k_d)\in\Z^{d-1}}k_1^2\\& \times\left(\left(\frac{k_1}{a}\right)^2+\sum_{j=2}^{d}
\left(\frac{k_j}{L_j}\right)^2\right)^{-1/2} \exp\left( -\frac{2\pi l}{T} \sqrt{\left(\frac{k_1}{a}\right)^2+\sum_{j=2}^{d}
\left(\frac{k_j}{L_j}\right)^2}\right)\nonumber\\&-\frac{\pi T^2}{6}
-2T\sum_{l=1}^{\infty}\sum_{(k_2,\ldots, k_d)\in\Z^{d-1}\setminus\{0\}} l^{-1} \sqrt{\sum_{j=2}^d \left(
\frac{k_j}{L_j}\right)^2}K_1\left(\frac{2\pi l}{T}\sqrt{\sum_{j=2}^d \left(
\frac{k_j}{L_j}\right)^2}\right)\nonumber
\end{align}for massless scalar field with Pb.c.; and by
\begin{align}\label{eq7_15_17}
&F_{\text{Cas}}  (a; L_2, \ldots, L_d) = F_{\text{Cas}}^{  T=0} (a; L_2, \ldots, L_d)+\frac{\pi}{a^3}
\sum_{l=1}^{\infty}\sum_{k_1=1}^{\infty}\sum_{(k_2,\ldots, k_d)\in(\mathbb{N}\cup\{0\})^{d-1}}k_1^2\\& \times\Lambda(k_2,\ldots, k_d)\left(\left(\frac{k_1}{a}\right)^2+\sum_{j=2}^{d}\nonumber
\left(\frac{k_j}{L_j}\right)^2\right)^{-1/2} \exp\left( -\frac{\pi l}{T} \sqrt{\left(\frac{k_1}{a}\right)^2+\sum_{j=2}^{d}
\left(\frac{k_j}{L_j}\right)^2}\right)\\&-\delta\frac{\pi T^2}{6}
-T\sum_{l=1}^{\infty}\sum_{(k_2,\ldots, k_d)\in(\mathbb{N}\cup\{0\})^{d-1}\setminus\{0\}} \Lambda(k_2,\ldots, k_d)l^{-1} \sqrt{\sum_{j=2}^d \left(
\frac{k_j}{L_j}\right)^2}K_1\left(\frac{\pi l}{T}\sqrt{\sum_{j=2}^d \left(
\frac{k_j}{L_j}\right)^2}\right)\nonumber
\end{align}for massless scalar field with Db.c.~ and Nb.c.~ and electromagnetic field with PEC b.c.~ and PMC b.c. Here
\begin{align*}
\Lambda^D(k_2,\ldots, k_d) = 1 \hspace{1cm}\text{if and only if all $k_i$, $i=2, \ldots, d$ are nonzero};
\end{align*} $\Lambda^N(k_2, \ldots, k_d) \equiv 1$ and $\Lambda^{PEC}(k_2,\ldots, k_d)$ and $\Lambda^{PMC}(k_2,\ldots, k_d)$
are given by \eqref{eq7_15_11} and \eqref{eq7_15_12} respectively; $\delta^N\equiv 1$; $\delta^{D }=\delta^{PMC}\equiv 0$ and $\delta^{PEC}=1$ if and only if $d=2$. These formulas show that for massless scalar field with Pb.c.~ and Nb.c., the temperature correction is of order $O(T^2)$ when $T\ll 1$; but for massless scalar field with Db.c.~ and for electromagnetic field with PEC b.c.~ and PMC b.c., the temperature correction terms go to zero exponentially fast when $T\rightarrow 0^+$.
The main contribution to the low temperature Casimir force comes from the first term of \eqref{eq7_15_16} or \eqref{eq7_15_17}, which is the zero temperature contribution. For massless scalar field with Pb.c., it is given by
\begin{align*}
&F_{\text{Cas}}^{P, T=0}(a; L_1, \ldots, L_d)= -\frac{d\Gamma\left(\frac{d+1}{2}\right)}{a^{d+1}\pi^{\frac{d+1}{2}}}\left[\prod_{j=2}^d L_j\right]\zeta_R(d+1)\\&-\frac{\Gamma\left(\frac{d+1}{2}\right)}{2\pi^{\frac{d+1}{2}}}\left[\prod_{j=2}^d L_j\right]
Z_{d-1}\left(\frac{d+1}{2}; L_2, \ldots, L_d\right)+\frac{4\pi\left[\prod_{j=2}^d L_j\right]}{a^{\frac{d+4}{2}}}\\&\times\sum_{k_1=1}^{\infty}
\sum_{(k_2,\ldots, k_d)\in \Z^{d-1}\setminus\{0\}}k_1^{\frac{d+2}{2}}\left(\sum_{j=2}^d \left(k_jL_j\right)^2\right)^{-\frac{d-2}{4}}
K_{\frac{d-2}{2}}\left( \frac{2\pi k_1}{a}\sqrt{\sum_{j=2}^d \left(k_jL_j\right)^2}\right).
\end{align*}The first term is the leading term and is of order $O(a^{-d-1})$ when $a\rightarrow 0^+$. We have seen in the previous section that for any finite temperature, this is still the leading term when $a\rightarrow 0^+$. The second term is $O(a^0)$ and the last term goes to zero exponentially fast when $a\rightarrow 0^+$. For massless scalar field with Db.c.~ and Nb.c.~ and electromagnetic field with PEC b.c.~ and PMC b.c., the leading behavior of the zero temperature Casimir force when $a\rightarrow 0^+$ is also the same as  the finite temperature case.

An alternative expression for the zero temperature Casimir force acting on the piston which can be used to study the large $a$ behavior is given by
\begin{align*}
&F_{\text{Cas}}^{P, T=0} (a; L_2, \ldots, L_d) \\=& -\frac{\pi}{6a^2} -\frac{2}{a} \sum_{k_1=1}^{\infty}\sum_{(k_2,\ldots, k_d)\in\Z^{d-1}\setminus\{0\}}k_1^{-1}\sqrt{\sum_{j=2}^d\left(\frac{k_j}{L_j}\right)^2}
K_1\left(2\pi k_1 a\sqrt{\sum_{j=2}^d\left(\frac{k_j}{L_j}\right)^2}\right)\\
&-4\pi \sum_{k_1=1}^{\infty}\sum_{(k_2,\ldots, k_d)\in\Z^{d-1}\setminus\{0\}}\left(\sum_{j=2}^d\left(\frac{k_j}{L_j}\right)^2\right)
K_0\left(2\pi k_1 a\sqrt{\sum_{j=2}^d\left(\frac{k_j}{L_j}\right)^2}\right)
\end{align*}for massless scalar field with Pb.c. For massless scalar field with Db.c.~ and Nb.c.~ and electromagnetic field with PEC b.c.~ and PMC b.c., \eqref{eq7_14_9}, \eqref{eq7_14_4} and \eqref{eq7_14_5} give immediately
\begin{align*}
&F_{\text{Cas}}^{ T=0} (a; L_2, \ldots, L_d) =-\delta \frac{\pi}{24 a^2} \\&-\frac{1}{2 a} \sum_{k_1=1}^{\infty}\sum_{(k_2,\ldots, k_d)\in\left(\mathbb{N}\cup\{0\}\right)^{d-1}\setminus\{0\}}\Lambda(k_2,\ldots, k_d)k_1^{-1}\sqrt{\sum_{j=2}^d\left(\frac{k_j}{L_j}\right)^2}
K_1\left(2\pi k_1 a\sqrt{\sum_{j=2}^d\left(\frac{k_j}{L_j}\right)^2}\right)\\
&-\pi \sum_{k_1=1}^{\infty}\sum_{(k_2,\ldots, k_d)\in\left(\mathbb{N}\cup\{0\}\right)^{d-1}\setminus\{0\}}\Lambda(k_2,\ldots, k_d)\left(\sum_{j=2}^d\left(\frac{k_j}{L_j}\right)^2\right)
K_0\left(2\pi k_1 a\sqrt{\sum_{j=2}^d\left(\frac{k_j}{L_j}\right)^2}\right).
\end{align*}Using the fact that
$$K_{\nu}(z) \sim\sqrt{\frac{\pi}{2z}}e^{-z}\hspace{1cm}\text{as}\;z\rightarrow \infty,$$ we find that when $a\gg 1$, the leading order terms of the zero temperature Casimir force are given respectively by
\begin{align*}
F_{\text{Cas}}^{P, T=0} (a; L_2, \ldots, L_d) \sim -\frac{\pi}{6a^2}-4\pi a^{-\frac{1}{2}}\left(
\min\{L_2^{-1}, \ldots, L_d^{-1}\}\right)^{\frac{3}{2}}e^{-2\pi a\min\{L_2^{-1}, \ldots, L_d^{-1}\} },
\end{align*}
\begin{align*}
F_{\text{Cas}}^{D, T=0} (a; L_2, \ldots, L_d) \sim-\pi a^{-\frac{1}{2}}\left(
\sum_{j=2}^d L_j^{-2}\right)^{\frac{3}{4}}\exp\left(-2\pi a\sqrt{\sum_{j=2}^d L_j^{-2}} \right),
\end{align*}\begin{align*}
F_{\text{Cas}}^{N, T=0}(a; L_2, \ldots, L_d) \sim &-\frac{\pi}{24a^2}-\pi a^{-\frac{1}{2}}\left(\min\{L_2^{-1},\ldots, L_d^{-1}\}\right)^{\frac{3}{2}} e^{-2\pi a\min\{L_2^{-1},
\ldots, L_d^{-1}\}},\end{align*}\begin{align*}
F_{\text{Cas}}^{PEC, T=0}(a; L_2, \ldots, L_d) \sim& -\delta\frac{\pi}{24a^2}- \pi a^{-\frac{1}{2}}\left(\min\left\{ \sum_{\substack{l=2\\l\neq j}}^dL_l^{-2}\right\}_{j=2}^d \right)^{\frac{3}{4}} \\&\times\exp\left(-2\pi a \min\left\{\sqrt{\sum_{\substack{l=2\\l\neq j}}^dL_l^{-2}}\right\}_{j=2}^d\right),\end{align*}\begin{align*}
F_{\text{Cas}}^{PMC}(a; L_2, \ldots, L_d) \sim &-\pi a^{-\frac{1}{2}}\left(\min\{L_2^{-1},\ldots, L_d^{-1}\}\right)^{\frac{3}{2}} e^{-2\pi a\min\{L_2^{-1},
\ldots, L_d^{-1}\}}.
\end{align*}We notice a considerable difference between the $a\rightarrow \infty$ leading behavior of the Casimir force in the zero temperature case and finite temperature case. At zero temperature, the leading order terms of the Casimir force for massless scalar field with Pb.c.~ and Nb.c.~ are of order $a^{-2}$, in contrast to the order $a^{-1}$ when the temperature is nonzero. For massless scalar field with Db.c.~ and for electromagnetic field with PEC b.c.~ and PMC b.c., the zero temperature Casimir force also tends to zero exponentially fast when $a\rightarrow \infty$; but it tends to zero faster than at finite temperature.

Now we consider the high temperature behavior of the Casimir force. From \eqref{eq7_16_4}, we find that for massless scalar field with Pb.c., the   contribution from region I to the Casimir force acting on the piston  is given by
\begin{align}\label{eq7_16_2}
&F_{\text{ Cas}}^{P, \text{I}}(a; L_2, \ldots, L_d) = \frac{\Gamma\left(\frac{d+1}{2}\right)\zeta_R(d+1)}{\pi^{\frac{d+1}{2}}}\left[\prod_{j=2}^d L_j\right]T^{d+1}
\\\nonumber&+\frac{T}{2}\frac{\pa}{\pa a}Z_d'\left(0; \frac{1}{a}, \frac{1}{L_2}, \ldots,\frac{1}{L_d}\right)+2\left[\prod_{j=2}^d L_j\right]T^{\frac{d+2}{2}}
\sum_{l=1}^{\infty}\sum_{\mathbf{k}\in\Z^d\setminus\{0\}}l^{\frac{d}{2}}\\&\times \left((k_1a)^2+\sum_{j=2}^d (k_jL_j)^2\right)^{-\frac{d}{4}}K_{\frac{d}{2}}
\left(2\pi l T\sqrt{(k_1a)^2+\sum_{j=2}^d (k_jL_j)^2}\right)\nonumber\\
&-8\pi a\left[\prod_{j=2}^d L_j\right]T^{\frac{d+4}{2}}
\sum_{l=1}^{\infty}\sum_{k_1=1}^{\infty}\sum_{(k_2,\ldots, k_d)\in \Z^{d-1}}l^{\frac{d+2}{2}}k_1^2\nonumber\\&\times \left((k_1a)^2+\sum_{j=2}^d (k_jL_j)^2\right)^{-\frac{d+2}{4}}K_{\frac{d+2}{2}}
\left(2\pi l T\sqrt{(k_1a)^2+\sum_{j=2}^d (k_jL_j)^2}\right).\nonumber
\end{align}The leading term is the force due to black body radiation and it is independent of $a$. For the   contribution from region II to the Casimir force acting on the piston, notice that $$F_{\text{ Cas}}^{P, \text{II}}(a; L_2, \ldots, L_d)=-\lim_{L_1\rightarrow \infty}
F_{\text{ Cas}}^{P, \text{I}}(L_1-a; L_2, \ldots, L_d)=-\lim_{a\rightarrow \infty}
F_{\text{ Cas}}^{P, \text{I}}( a; L_2, \ldots, L_d).$$ Therefore, we find that the black body radiation contributions from   regions I and II cancel with each other, and the  Casimir force acting on the piston is given by
\begin{align}\label{eq7_16_3}
&F_{\text{ Cas}}^{P}(a; L_2, \ldots, L_d) =\frac{T}{2}\frac{\pa}{\pa a}Z_d'\left(0; \frac{1}{a}, \frac{1}{L_2}, \ldots,\frac{1}{L_d}\right)\\&
-\frac{T}{2}\lim_{a\rightarrow \infty}\frac{\pa}{\pa a}Z_d'\left(0; \frac{1}{a}, \frac{1}{L_2}, \ldots,\frac{1}{L_d}\right)+4\left[\prod_{j=2}^d L_j\right]T^{\frac{d+2}{2}}\nonumber
\sum_{l=1}^{\infty}\sum_{k_1=1}^{\infty}\sum_{(k_2,\ldots, k_d)\in\Z^{d-1}}l^{\frac{d}{2}}\\\nonumber&\times \left((k_1a)^2+\sum_{j=2}^d (k_jL_j)^2\right)^{-\frac{d}{4}}K_{\frac{d}{2}}
\left(2\pi l T\sqrt{(k_1a)^2+\sum_{j=2}^d (k_jL_j)^2}\right)\end{align}\begin{align*}&-8\pi a^2\left[\prod_{j=2}^d L_j\right]T^{\frac{d+4}{2}}
\sum_{l=1}^{\infty}\sum_{k_1=1}^{\infty}\sum_{(k_2,\ldots, k_d)\in \Z^{d-1}}l^{\frac{d+2}{2}}k_1^2\nonumber\\&\times \left((k_1a)^2+\sum_{j=2}^d (k_jL_j)^2\right)^{-\frac{d+2}{4}}K_{\frac{d+2}{2}}
\left(2\pi l T\sqrt{(k_1a)^2+\sum_{j=2}^d (k_jL_j)^2}\right).\nonumber
\end{align*}Notice that the high temperature leading term is linear in $T$, given by the sum of the first two terms. It is called the \emph{classical term} of the Casimir force. The remaining (last two) terms decays exponentially when $T\rightarrow \infty$.

Since the sign of the Stefan--Boltzmann term in \eqref{eq7_16_2} is positive, the contribution to the Casimir force from   region I will become very repulsive at high temperature. However, the black body radiation of   region II creates a counter force to cancel this repulsive force  and we find that \emph{the leading order term of the Casimir force acting on the piston  at high temperature is an attractive force} (see \eqref{eq7_16_5}) \emph{of order $T$}.

Now let us compute explicitly the classical term of the Casimir force for massless scalar field with Pb.c. Using  \eqref{eq6_12_15}, we find that   the classical term of the Casimir force acting on the piston for massless scalar field with Pb.c.~ is
\begin{align}\label{eq7_16_5}
&F_{\text{ Cas}}^{P, \text{classical}}(a; L_2, \ldots, L_d)\\= &-T\left(\frac{1}{a}+2\pi \sum_{k_1=1}^{\infty}\sum_{(k_2, \ldots, k_d)\in\Z^{d-1}\setminus\{0\}}
\sqrt{\sum_{j=2}^d\left(\frac{k_j}{L_j}\right)^2}\exp\left(-2\pi k_1 a\sqrt{\sum_{j=2}^d\left(\frac{k_j}{L_j}\right)^2}\right)\right).\nonumber
\end{align}
A drawback with the formula \eqref{eq7_16_5}  is that it does not manifest the small $a$ behavior of the classical term. Using   \eqref{eq6_12_15} again, we obtain an alternative formula  for the classical term
\begin{align}\label{eq7_18_2}
&F_{\text{Cas}}^{P, \text{classical}}(a; L_2, \ldots, L_d)= -T\Biggl\{ \frac{(d-1)}{\pi^{\frac{d}{2}}}\Gamma\left(\frac{d}{2}\right)\zeta_R(d) \left[\prod_{j=2}^d L_j\right]a^{-d} -4\pi a^{-\frac{d+3}{2}}\\&\times \left[\prod_{j=2}^d L_j\right]\sum_{k_1=1}^{\infty}  \sum_{(k_2, \ldots, k_d)\in\Z^{d-1}\setminus\{0\}}\left(
\sum_{j=2}^d (k_jL_j)^2\right)^{-\frac{d-3}{4}}k_1^{\frac{d+1}{2}}K_{\frac{d-3}{2}}\left(\frac{2\pi k_1}{a}\sqrt{\sum_{j=2}^d (k_jL_j)^2}
\right)\nonumber\\
&+\frac{\Gamma\left(\frac{d}{2}\right)}{2\pi^{\frac{d}{2}}} \left[\prod_{j=2}^d L_j\right]Z_{d-1}\left(\frac{d}{2}; L_2,\ldots, L_d\right)\Biggr\}\nonumber
\end{align}for massless scalar field with Pb.c., which is suitable for studying small $a$ behavior.

The high temperature expansion  for Casimir force acting on the piston due to massless scalar field with Db.c.~ and Nb.c.~ and electromagnetic field with PEC b.c.~ and PMC b.c.~ can be obtained using \eqref{eq7_14_9}, \eqref{eq7_14_4} and \eqref{eq7_14_5}. For the   contribution from region I, \eqref{eq7_14_9}, \eqref{eq7_14_4}, \eqref{eq7_14_5} and \eqref{eq7_16_2} show that leading order terms of the Casimir force from   region I  are given by
\begin{align*}
F_{\text{Cas}}^{D/N,\text{I}}(a; L_2, \ldots, L_d) = \sum_{j=1}^{d}\left[(\mp)^{d-j}\frac{\Gamma\left(\frac{j+1}{2}\right)\zeta_R(j+1)}{2^{d-j}\pi^{\frac{j+1}{2}}}S_j\right] T^{j+1}+O(T)
\end{align*}and
\begin{align*}
F_{\text{Cas}}^{PEC/PMC,\text{I}}(a; L_2, \ldots, L_d) = \sum_{j=1}^{d}\left[(\mp)^{d-j}(2j-d-1)\frac{\Gamma\left(\frac{j+1}{2}\right)\zeta_R(j+1)}{2^{d-j}\pi^{\frac{j+1}{2}}}S_j\right] T^{j+1}+O(T)
\end{align*}respectively, where $S_j$ is the partial hypersurface area defined by \eqref{eq7_25_1}. Notice that in contrast to the Pb.c.~ case, now we have temperature corrections of order $T^{l}$ for all $2\leq l\leq d$ besides the leading Stefan--Boltzmann term of order $T^{d+1}$. However, all these terms are independent of $a$ and therefore they cancel with the corresponding terms from   region II. The high temperature leading order term of the Casimir force acting on the piston is still the classical term of order $T$ and is given explicitly by
\begin{align}\label{eq7_16_6}
&F_{\text{ Cas}}^{\text{classical}} (a; L_2, \ldots, L_d)=-T\Biggl(\frac{\delta}{2a}+\pi \sum_{k_1=1}^{\infty}\sum_{(k_2, \ldots, k_d)\in
\left(\mathbb{N}\cup\{0\}\right)^{d-1}\setminus\{0\}}\\&\Lambda(k_2, \ldots, k_d)\sqrt{\sum_{j=2}^d\left(\frac{k_j}{L_j}\right)^2}
\exp\left(-2\pi k_1 a\sqrt{\sum_{j=2}^d\left(\frac{k_j}{L_j}\right)^2}\right)\Biggr).\nonumber
\end{align}
By taking the derivative of \eqref{eq7_16_5} and \eqref{eq7_16_6} with respect to $a$, we find that the classical term of the Casimir force is also an increasing function of $a$. In other words, the magnitude of the classical term decreases with increasing $a$. An alternative expression suitable for studying the classical term of the Casimir force at small plate separation can be derived using \eqref{eq7_14_9}, \eqref{eq7_14_4}, \eqref{eq7_14_5} and \eqref{eq7_18_2}.

\begin{figure}\centering \epsfxsize=.49\linewidth
\epsffile{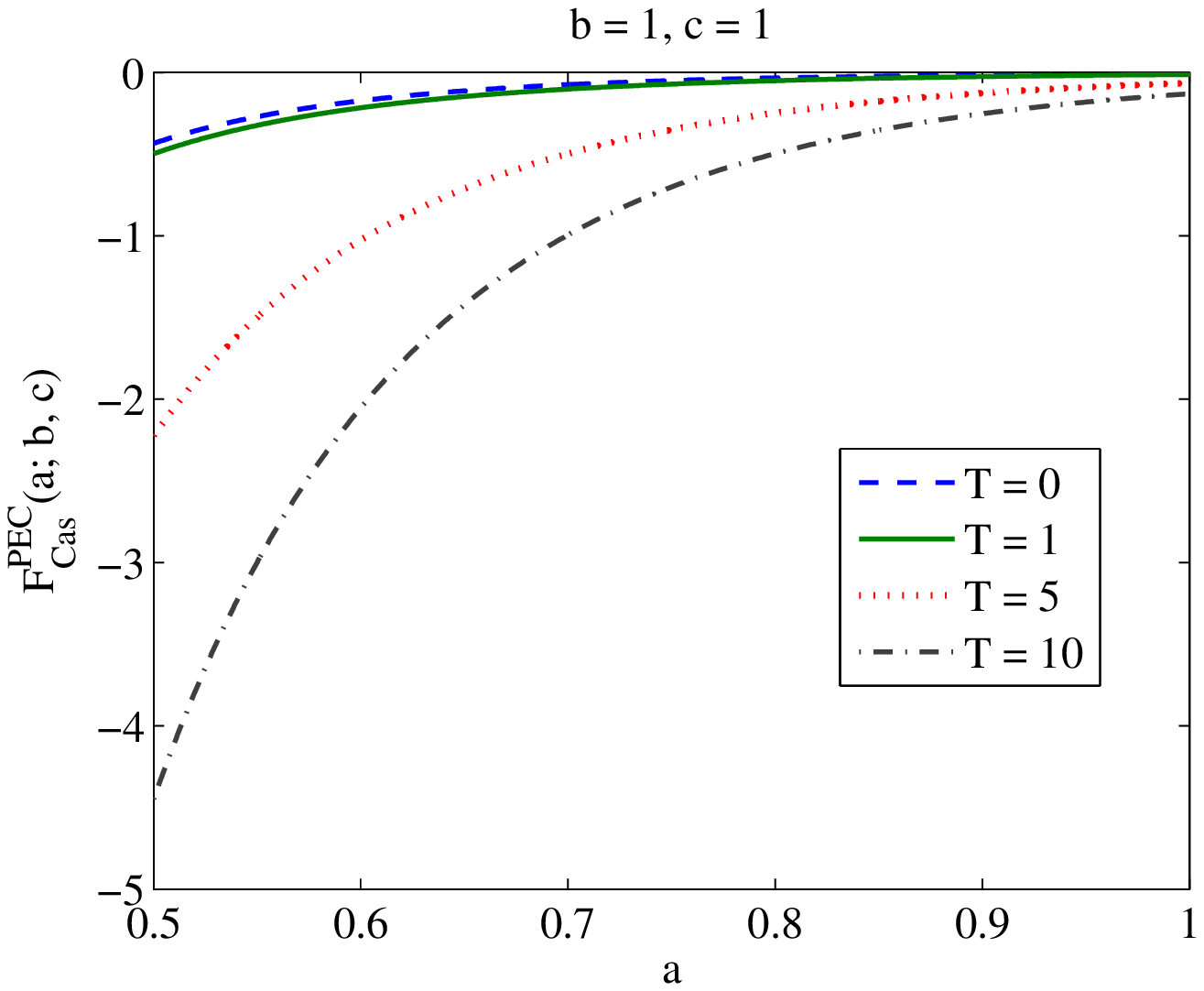} \epsfxsize=.49\linewidth
\epsffile{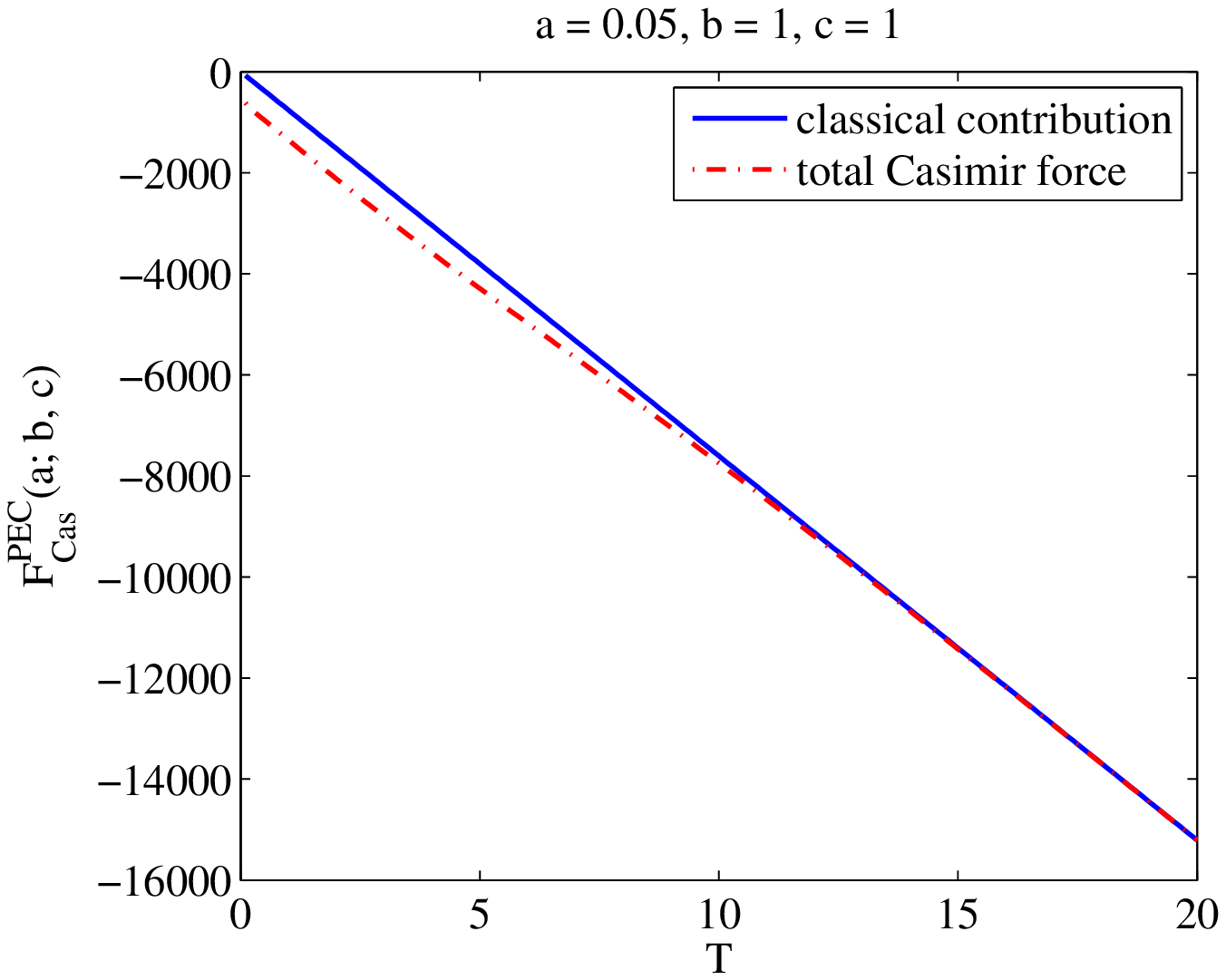}\caption{Left: The dependence of the Casimir force $F_{\text{Cas}}^{PEC}(a; b,c)$ on the plate separation $a$ when $T=0, 1, 5, 10$. Right: The deviation of the Casimir force from the classical term when $a=0.05, b=1, c=1$.}\end{figure}

We would like to remark that the thermal Casimir effect for electromagnetic field in a three dimensional rectangular cavity is considered in \cite{44}. In that paper, the authors obtain Casimir force that depends linearly in $T$ in the high temperature regime by subtracting the terms \begin{align}\label{eq8_29_1}\sum_{j=1}^{d}\left[(\mp)^{d-j}(2j-d-1)\frac{\Gamma\left(\frac{j+1}{2}\right)\zeta_R(j+1)}{2^{d-j}\pi^{\frac{j+1}{2}}}S_j\right] T^{j+1}\end{align} so that in the large volume limit, the Casimir force tends to zero. In our approach, no subtraction is required since the contribution from the terms \eqref{eq8_29_1} automatically cancel.

Before ending this section, we would like  to comment on the nomenclature 'classical term' for the high temperature leading term. To restore the constants $\hbar, c$ and $k$ into the expressions for the Casimir force, we replace $T$ by $kT/(\hbar c)$ everywhere and multiply the overall expression by $\hbar c$. Notice that a term with order $T^j$ will be accompanied with $\hbar^{1-j}$. Since in the high temperature limit, the Casimir force acting on the piston is $O(T)$, this implies that it has a finite classical  $\hbar\rightarrow 0^+$ limit. Moreover, the leading term linear in $T$ gives the classical limit and therefore it is called the classical term.

In Appendix B, we compute the explicit expressions for the zero temperature Casimir force when $d=1, 2$ and $3$ from the formulas in this section. The results are found to agree with the existing results in \cite{14, 15, 16, 25, 26, 29}. Explicit formulas for  the low and high temperature expansions of the Casimir force are also given. In Figure 2, we show the dependence of the Casimir force $F_{\text{Cas}}^{PEC}(a; b,c)$ on the plate separation $a$ and temperature $T$ for three dimensional electromagnetic field with PEC b.c.

\section{Comment on piston inside a closed cavity}

Consider the case where the piston is confined in a closed rectangular cavity. More precisely, we consider the scenario  where the $L_1\rightarrow \infty$ limit is not taken.   In showing that the Casimir force is divergence free in the beginning of Section 3, we only use the fact that the divergent part of the Casimir energy in  a rectangular cavity depends linearly on $L_1$ (without passing to the limit $L_1\rightarrow \infty$). Therefore in the present case where $L_1$ is finite, it is still true that the Casimir force acting on the piston $\hat{F}_{\text{Cas}}(a; L_1, \ldots, L_d)$ is free of divergence. On the other hand,  the derivation of \eqref{eq7_14_1} from \eqref{eq7_22_3} shows that after some cancelations, the Casimir force $\hat{F}_{\text{Cas}}(a; L_1, \ldots, L_d)$ for finite $L_1$ case can be written in terms of the Casimir force $F_{\text{Cas}}(a; L_2,\ldots, L_d)$ of infinite $L_1$ case by
\begin{align}\label{eq7_22_4}
\hat{F}_{\text{Cas}}(a; L_1, \ldots, L_d)=F_{\text{Cas}}(a; L_2,\ldots, L_d) - F_{\text{Cas}}(L_1-a; L_2,\ldots, L_d).
\end{align}This is true for either massless scalar field with Pb.c., Db.c.~ or Nb.c.~ or electromagnetic field with PEC b.c.~ or PMC b.c. A trivial consequence of \eqref{eq7_22_4} is that the Casimir force acting on a piston placed in a closed rectangular cavity is equal to zero when the piston is  placed exactly in the middle of the cavity. On the other hand, since we have shown that in all the cases we consider,    $F_{\text{Cas}}(a; L_2, \ldots, L_d)$ is an increasing function of $a$, this implies that if the piston is placed closer to the left hand side (i.e. $a< L_1-a$), then the Casimir force on the piston $\hat{F}_{\text{Cas}}(a; L_1, \ldots, L_d)$ is negative and tends to pull it to the left. In other words, the Casimir force acting on a piston which is placed inside a closed rectangular cavity always try to collapse the piston to the nearer end.

The low and high temperature expansions of the Casimir force $\hat{F}_{\text{Cas}}(a; L_1, \ldots, L_d)$ in the case of finite $L_1$ can  easily be computed using the formula \eqref{eq7_22_4} and the results for the case of infinite $L_1$ in the previous sections. Therefore we omit them here.

\section{ Casimir force density when the distances of some pairs of transversal plates are large}
In this section, we   want to study the asymptotic behavior of the Casimir force when $d-p >0$   pairs of the transversal plates are large, i.e., $a, L_1, \ldots, L_p \ll L_{p+1}= \ldots= L_d$ for some $p\geq 1$. More precisely, we are going to derive the limit
\begin{align*}
\mathcal{F}_{\text{Cas}}(p,d; a; L_2, \ldots, L_p) =\lim_{L_{p+1}=\ldots=L_d\rightarrow \infty} \frac{F_{\text{Cas}}(a; L_1, \ldots, L_d)}{L_{p+1}\ldots L_d},
\end{align*}which we call the Casimir force density\footnote{When $p=1$, this is actually the pressure on the piston when the cross section of the piston becomes an infinite hyperplane.}.
We consider the low temperature expansion and high temperature expansion separately. The computations are similar as in the previous sections. We only write down the final answers. Details can be found in the preprint version of this paper posted in the arXiv \cite{45}.

\subsection{Low temperature expansion}
In the low temperature regime, the Casimir force density acting on the piston is given by
\begin{align*}
&\mathcal{F}_{\text{Cas}}^{P}(p, d; a, L_2, \ldots, L_d) = \mathcal{F}_{\text{Cas}}^{P, T=0}(p, d; a, L_2, \ldots, L_d) -2T^{\frac{d-p+2}{2}}\\&\times\sum_{(k_2, \ldots, k_p)\in\Z^{p-1}\setminus\{0\}} \sum_{l=1}^{\infty}\left( \sum_{j=2}^p \left(\frac{k_j}{L_j}\right)^2\right)^{\frac{d-p+2}{4}}l^{-\frac{d-p+2}{2}} K_{\frac{d-p+2}{2}}\left( \frac{2\pi l}{T} \sqrt{ \sum_{j=2}^p \left(\frac{k_j}{L_j}\right)^2}\right)\\&+\frac{8\pi}{a^3}T^{\frac{d-p-1}{2}}\sum_{k_1=1}^{\infty} \sum_{(k_2, \ldots, k_p)\in\Z^{p-1}}\sum_{l=1}^{\infty}\left( \left(\frac{k_1}{a}\right)^2+\sum_{j=2}^p \left(\frac{k_j}{L_j}\right)^2\right)^{\frac{d-p-1}{4}}l^{-\frac{d-p-1}{2}}k_1^2\\&\times K_{\frac{d-p-1}{2}}\left( \frac{2\pi l}{T} \sqrt{ \left(\frac{k_1}{a}\right)^2+\sum_{j=2}^p \left(\frac{k_j}{L_j}\right)^2
}\right)-\frac{\Gamma\left(\frac{d-p+2}{2}\right)}{\pi^{\frac{d-p+2}{2}}}T^{d-p+2}\zeta_R(d-p+2)
\end{align*}for massless scalar field with Pb.c. The expressions of the zero temperature Casimir force density which are suitable for studying the small $a$ and large $a$ behaviors are given by
\begin{align*}
&\mathcal{F}_{\text{Cas}}^{P, T=0}(p, d; a, L_2, \ldots, L_d) =-\frac{d
\Gamma\left(\frac{d+1}{2}\right)}{\pi^{\frac{d+1}{2}}}\left[\prod_{j=2}^p L_j\right]a^{-d-1}-\frac{
\Gamma\left(\frac{d+1}{2}\right)}{2\pi^{\frac{d+1}{2}}}\left[\prod_{j=2}^p L_j\right]\\&\times Z_{E, p-1}\left(\frac{d+1}{2}; L_2,\ldots, L_p\right) +\frac{4\pi}{a^{\frac{d+4}{2}}}\left[\prod_{j=2}^p L_j\right]\sum_{k_1=1}^{\infty}
\sum_{(k_2,\ldots, k_p)\in\Z^{p-1}\setminus\{0\}} k_1^{\frac{d+2}{2}}\\
&\times\left(\sum_{j=2}^p (k_jL_j)^2\right)^{-\frac{d-2}{4}} K_{\frac{d-2}{2}}\left(\frac{2\pi k_1}{a}\sqrt{\sum_{j=2}^p (k_jL_j)^2}\right);
\end{align*}and
\begin{align*}
&\mathcal{F}_{\text{Cas}}^{P, T=0}(p, d; a, L_2, \ldots, L_d) =
-\frac{d-p+1}{\pi^{\frac{d-p+2}{2}}}\Gamma\left(\frac{d-p+2}{2}\right)\zeta_R(d-p+2)a^{-d+p-2}\\
&-2(d-p+1)a^{-\frac{d-p+2}{2}}\sum_{k_1=1}^{\infty}\sum_{(k_2, \ldots, k_p)\in\Z^{p-1}\setminus\{0\}}\left(\sum_{j=2}^p \left(\frac{k_j}{L_j}\right)^2
\right)^{\frac{d-p+2}{4}}k_1^{-\frac{d-p+2}{2}}\\
&\times K_{\frac{d-p+2}{2}}\left(2\pi k_1 a\sqrt{\sum_{j=2}^p \left(\frac{k_j}{L_j}\right)^2}
\right)-4\pi a^{-\frac{d-p}{2}}\sum_{k_1=1}^{\infty}\sum_{(k_2, \ldots, k_p)\in\Z^{p-1}\setminus\{0\}}\left(\sum_{j=2}^p \left(\frac{k_j}{L_j}\right)^2
\right)^{\frac{d-p+4}{4}}\\
&\times k_1^{-\frac{d-p}{2}} K_{\frac{d-p}{2}}\left(2\pi k_1 a\sqrt{\sum_{j=2}^p \left(\frac{k_j}{L_j}\right)^2}
\right)
\end{align*} respectively. For massless scalar field with Db.c.~ and Nb.c.~ and electromagnetic field with PEC b.c.~ and PMC b.c., the results can be obtained from that for massless scalar field with Pb.c.~ by using
\begin{align}\label{eq7_17_7}&\mathcal{F}_{\text{Cas}}^{D/N}(p, d; a; L_2, \ldots, L_p) \\=&
2^{-p+1}\sum_{j=1}^p (\mp)^{p-j}\sum_{2\leq m_1<\ldots < m_{j-1}\leq p}\mathcal{F}_{\text{Cas}}^{P}(j, j+d-p; 2a; 2L_{m_1}, \ldots, 2L_{m_{j-1}}); \nonumber \end{align}
\begin{align}\label{eq7_17_8}
&\mathcal{F}_{\text{Cas}}^{PEC}(p, d; a; L_2, \ldots, L_p) = (d-1)\mathcal{F}_{\text{Cas}}^{D}(p, d; a; L_2, \ldots, L_p)\\&\hspace{4cm}+\sum_{j=2}^{p}\mathcal{F}_{\text{Cas}}^{D}(p-1, d-1; a; L_2, \ldots, L_{j-1}, L_{j+1},\ldots, L_p)\nonumber; \end{align}
\begin{align}\label{eq7_17_9}&\mathcal{F}_{\text{Cas}}^{PMC}(p, d; a; L_2, \ldots, L_p)\\
=&\sum_{j=1}^p (d-p+j-1)\sum_{2\leq m_1<\ldots < m_j\leq p}\mathcal{F}_{\text{Cas}}^{D}(j, j+d-p; a; L_{m_1}, \ldots, L_{m_j});\nonumber\end{align}
which can be derived from \eqref{eq7_14_9}, \eqref{eq7_14_4} and \eqref{eq7_14_5} (see \cite{39}). We notice that when $p\geq 2$,  the  leading order terms of the temperature correction to the Casimir force density is of order $T^{d-p+2}$ when $T\ll 1$ for massless scalar field with Pb.c.~ and Nb.c. However, for massless scalar field with Db.c.~ and electromagnetic field with PEC b.c.~ and PMC b.c., the temperature correction to the Casimir force density decays to zero exponentially fast when $T\rightarrow 0$. The main contribution to the low temperature Casimir force density comes from the zero temperature Casimir force density which at small plate separation $a$, has leading order   proportional to $-a^{-d-1}$ that goes to  negative infinity. At large $a$, the leading order term is proportional to $-a^{-d+p-2}$ for massless scalar field with Pb.c.~ and Nb.c.; and decays exponentially for massless scalar field with Db.c.~ and electromagnetic field with PEC b.c.~ and PMC b.c.

In the particular case where $p=1$, i.e. when the piston becomes a pair of infinite parallel hyperplanes, we find that the low temperature expansion of the Casimir pressure acting on the piston is given by
\begin{align*}
\mathcal{P}_{\text{Cas}}^P(d; a) =&-\frac{d\Gamma\left(\frac{d+1}{2}\right)}{\pi^{\frac{d+1}{2}}}\zeta_R(d+1)a^{-d-1}
-\frac{\Gamma\left(\frac{d+1}{2}\right)}{\pi^{\frac{d+1}{2}}}T^{d+1}\zeta_R(d+1)\\&+\frac{8\pi}{a^{\frac{d+4}{2}}}
T^{\frac{d-2}{2}}\sum_{k=1}^{\infty}\sum_{l=1}^{\infty}k^{\frac{d+2}{2}}l^{-\frac{d-2}{2}}K_{\frac{d-2}{2}}\left(\frac{2\pi l k}{T a}\right)
\end{align*}for massless scalar field with Pb.c. For massless scalar field with Db.c.~ and Nb.c., the Casimir pressure is equal to $\mathcal{P}_{\text{Cas}}^P(d; 2a)$, i.e.
\begin{align*}\mathcal{P}_{\text{Cas}}^{D/N}(d; a) =&-\frac{d\Gamma\left(\frac{d+1}{2}\right)}{(4\pi)^{\frac{d+1}{2}}}\zeta_R(d+1)a^{-d-1}
-\frac{\Gamma\left(\frac{d+1}{2}\right)}{\pi^{\frac{d+1}{2}}}T^{d+1}\zeta_R(d+1)
\\&+\frac{\pi}{2^{\frac{d-2}{2}}a^{\frac{d+4}{2}}}
T^{\frac{d-2}{2}}\sum_{k=1}^{\infty}\sum_{l=1}^{\infty}k^{\frac{d+2}{2}}l^{-\frac{d-2}{2}}K_{\frac{d-2}{2}}\left(\frac{\pi l k}{T a}\right).\end{align*}For electromagnetic field with PEC b.c.~ and PMC b.c., the Casimir pressure is $(d-1)$ times larger than that of massless scalar field with Db.c., i.e. \begin{align*}\mathcal{P}_{\text{Cas}}^{PEC/PMC}(d; a) =(d-1)\mathcal{P}_{\text{Cas}}^{D/N}(d; a). \end{align*}
 Notice that the zero temperature Casimir pressure on a pair of infinite parallel plates is equal to
\begin{align*}
\mathcal{P}_{\text{Cas}}^{P, T=0}(d; a) =&-\frac{d\Gamma\left(\frac{d+1}{2}\right)}{\pi^{\frac{d+1}{2}}}\zeta_R(d+1)a^{-d-1}
\end{align*}for massless scalar field with Pb.c. It is $2^{-d-1}$ times weaker for massless scalar field with Db.c.~ and Nb.c., i.e. $\mathcal{P}_{\text{Cas}}^{D/N, T=0}(d; a)=2^{-d-1}\mathcal{P}_{\text{Cas}}^{P, T=0}(d; a)$. For electromagnetic field with PEC b.c.~ or PMC b.c., the zero temperature Casimir pressure on a pair of infinite parallel plates is
\begin{align*}
\mathcal{P}_{\text{Cas}}^{PEC/PMC}(d; a) = -\frac{d(d-1)}{(2\pi)^{\frac{d+1}{2}}}\Gamma\left(\frac{d+1}{2}\right)\zeta_R(d+1)a^{-d-1}.
\end{align*}These agree with the well known results.
For  the temperature correction, we find that the leading term is of order $T^{d+1}$ when $T\ll 1$. In particular, the leading thermal correction to Casimir force for massless scalar field with Dirichlet boundary condition is $$-\frac{\Gamma\left(\frac{d+1}{2}\right)}{\pi^{\frac{d+1}{2}}}T^{d+1}\zeta_R(d+1),$$ in agreement with the result in \cite{24_3}.

\subsection{High temperature expansion}
In the high temperature regime, we obtain  from \eqref{eq7_16_3} the following expansion for the Casimir force density for massless scalar field with Pb.c.:\begin{align*}
&\mathcal{F}_{\text{Cas}}^P(p, d; a; L_2, \ldots, L_p)= B^P(a; L_2, \ldots, L_p) T +4\left[\prod_{j=2}^p L_j\right]T^{\frac{d+2}{2}}\nonumber
\sum_{l=1}^{\infty}\sum_{k_1=1}^{\infty}\sum_{(k_2,\ldots, k_p)\in\Z^{p-1}}l^{\frac{d}{2}}\\\nonumber&\times \left((k_1a)^2+\sum_{j=2}^p (k_jL_j)^2\right)^{-\frac{d}{4}}K_{\frac{d}{2}}
\left(2\pi l T\sqrt{(k_1a)^2+\sum_{j=2}^p (k_jL_j)^2}\right)\end{align*}\begin{align*}&-8\pi a^2\left[\prod_{j=2}^p L_j\right]T^{\frac{d+4}{2}}
\sum_{l=1}^{\infty}\sum_{k_1=1}^{\infty}\sum_{(k_2,\ldots, k_p)\in \Z^{p-1}}l^{\frac{d+2}{2}}k_1^2\nonumber\\&\times \left((k_1a)^2+\sum_{j=2}^p (k_jL_j)^2\right)^{-\frac{d+2}{4}}K_{\frac{d+2}{2}}
\left(2\pi l T\sqrt{(k_1a)^2+\sum_{j=2}^p (k_jL_j)^2}\right).
\end{align*}In the high temperature limit, the last two terms decays exponentially while the leading term is the classical term linear in $T$ with coefficient $B^P(a; L_2, \ldots, L_p)$ which can be computed from \eqref{eq7_18_2}. Using   \eqref{eq7_15_18},
one gets\begin{align*}
&B^P(a; L_2, \ldots, L_p) = -\frac{(d-1)}{\pi^{\frac{d}{2}}}\Gamma\left(\frac{d}{2}\right)\zeta_R(d) \left[\prod_{j=2}^p L_j\right]a^{-d} + 4\pi a^{-\frac{d+3}{2}}\\&\times \left[\prod_{j=2}^p L_j\right]\sum_{k_1=1}^{\infty}  \sum_{(k_2, \ldots, k_p)\in\Z^{p-1}\setminus\{0\}}\left(
\sum_{j=2}^p (k_jL_j)^2\right)^{-\frac{d-3}{4}}k_1^{\frac{d+1}{2}}K_{\frac{d-3}{2}}\left(\frac{2\pi k_1}{a}\sqrt{\sum_{j=2}^p (k_jL_j)^2}
\right)\nonumber\\
&-\frac{\Gamma\left(\frac{d}{2}\right)}{2\pi^{\frac{d}{2}}}\left[\prod_{j=2}^pL_j\right]Z_{p-1}\left(\frac{d}{2}; L_2, \ldots, L_p\right).
\end{align*}The corresponding results for massless scalar field with Db.c.~ and Nb.c.~ and electromagnetic field with PEC b.c.~ and PMC b.c.~ can be obtained using \eqref{eq7_17_7}, \eqref{eq7_17_8} and \eqref{eq7_17_9}. Again, we find that in the high temperature limit, the leading order term of the  Casimir force density is the classical term of order $T$. It can be very negative when the plate separation $a$ is small.

Consider the particular case  where $p=1$. The high temperature expansion of the Casimir pressure for the massless scalar field with Pb.c.~ is given by
\begin{align*}
&\mathcal{P}_{\text{Cas}}^P(d; a) = -\frac{(d-1)}{\pi^{\frac{d}{2}}}\Gamma\left(\frac{d}{2}\right)\zeta_R(d) a^{-d}T+4T^{\frac{d+2}{2}}a^{-\frac{d}{2}}\nonumber
\sum_{l=1}^{\infty}\sum_{k =1}^{\infty}l^{\frac{d}{2}}k^{-\frac{d}{2}}\\\nonumber&\times K_{\frac{d}{2}}
\left(2\pi l kTa\right)-8\pi a^{-\frac{d-2}{2}} T^{\frac{d+4}{2}}
\sum_{l=1}^{\infty}\sum_{k=1}^{\infty}l^{\frac{d+2}{2}}k^{-\frac{d-2}{2}} K_{\frac{d+2}{2}}
\left(2\pi l k Ta\right).
\end{align*}For massless scalar field with Db.c.~ and Nb.c., it is given by
\begin{align}\label{eq7_18_3}
&\mathcal{P}_{\text{Cas}}^{D/N}(d; a) = -\frac{(d-1)}{2^d\pi^{\frac{d}{2}}}\Gamma\left(\frac{d}{2}\right)\zeta_R(d) a^{-d}T+2^{-\frac{d-4}{2}}T^{\frac{d+2}{2}}a^{-\frac{d}{2}}
\sum_{l=1}^{\infty}\sum_{k =1}^{\infty}l^{\frac{d}{2}}k^{-\frac{d}{2}}\\\nonumber&\times K_{\frac{d}{2}}
\left(4\pi l kTa\right)-2^{-\frac{d-8}{2}}\pi a^{-\frac{d-2}{2}} T^{\frac{d+4}{2}}
\sum_{l=1}^{\infty}\sum_{k=1}^{\infty}l^{\frac{d+2}{2}}k^{-\frac{d-2}{2}} K_{\frac{d+2}{2}}
\left(4\pi l k Ta\right).\nonumber
\end{align}For electromagnetic field with PEC b.c.~ and PMC b.c., it is $(d-1)$ times of \eqref{eq7_18_3}. The leading term in \eqref{eq7_18_3} agrees with those given in \cite{24_3}.

\section{Conclusion}

   We   consider finite temperature massless scalar field with Pb.c., Db.c.~ and Nb.c.~ and electromagnetic field with PEC b.c.~ and PMC b.c.~ for any space dimension $d$. For a rectangular piston which can be considered as a one--sided open rectangular cavity divided into two regions, it is shown that one can obtain a finite unambiguous Casimir force acting on the piston. Different exact expressions of the Casimir force which are suitable for studying the small and large plate separation limits  and low and high temperature limits  are derived. It is verified analytically that for all the cases we considered, although the regularized Casimir force acting on a wall of a rectangular cavity can be attractive or repulsive depending on the relative size of the cavity,  the Casimir force acting on the piston is always an attractive force. Moreover, the magnitude of the Casimir force decreases as the separation distance between the piston and the opposite wall increases. Another interesting result obtained in this paper is that  at high temperature,  the magnitude of the Casimir force is found to grow linearly in temperature $T$. This is in contrast to the result for rectangular cavities, where   at high temperature the leading term of the Casimir force is the Stefan--Boltzmann term of order $T^{d+1}$, an order much larger than $T$. It also shows that the Casimir force has a classical $\hbar\rightarrow 0$ limit.  On the other hand, we also establish that at low temperature,  the effect of temperature to the magnitude of the Casimir force is insignificant when the plate separation is small.

   We have derived   exact expressions for the Casimir force acting on the piston which are suitable for studying low and high temperature, and small and large plate separation limits. A more detailed numerical study of the results would be considered in a future work.
The methods used in this paper can be easily extended to other quantum fields as well as pistons with arbitrary cross section. In particular,  it will be interesting to consider massive field, fermionic field or massless field with mixed boundary conditions. The later will be a possible candidate for  repulsive Casimir force. Another possible direction is to study the Casimir piston  made of dielectric and magnetic materials.

 \vspace{1cm} \noindent \textbf{Acknowledgement}\;
This project is   supported by the Scientific Advancement Fund Allocation (SAGA) Ref. No P96c and e-Science fund 06-02-01-SF0080.

\appendix
\section{Formulas for Epstein zeta function and Casimir energy}
Here we gather the formulas we need for computing the Epstein zeta function \eqref{eq10_8_3}. The Chowla--Selberg formula \cite{4,5,37,38} for Epstein zeta function says that\begin{equation}\label{eq7_15_18}\begin{split}
Z_d(s; a_1, \ldots, a_d) =& Z_p(s; a_1 \ldots, a_p) + \frac{\pi^{p/2}\Gamma\left(s-\frac{p}{2}\right)}{\left[\prod_{j=1}^{p} a_j \right]\Gamma(s)}Z_{d-p}\left(s-\frac{p}{2}; a_{p+1}, \ldots, a_d\right)\\
&+ \frac{2\pi^{s}}{\left[\prod_{j=1}^{p} a_j \right]\Gamma(s)}\sum_{\mathbf{k}\in (\Z^p\setminus\{0\})\times
(\Z^{d-p}\setminus\{0\})}\left( \frac{\sum_{j=1}^p \left(\frac{k_j}{a_j}\right)^2}{\sum_{j=p+1}^d (k_j a_j)^2}\right)^{\frac{2s-p}{4}}\\
&\times K_{s-\frac{p}{2}}\left( 2\pi\sqrt{ \left(\sum_{j=1}^p \left(\frac{k_j}{a_j}\right)^2\right)\left(\sum_{j=p+1}^d (k_j a_j)^2\right)}\right).
\end{split}\end{equation}Taking derivative at $s=0$, we have
\begin{equation}\label{eq6_12_15}\begin{split}
&Z_{d+1}'(0;a_1,\ldots, a_d)
= Z_{p}'(0; a_1, \ldots,
a_p)+\frac{\pi^{\frac{p}{2}}\Gamma\left(-\frac{p}{2}\right)}{\left[\prod_{j=1}^pa_j\right]}Z_{d+1-p}\left(-\frac{p}{2};
a_{p+1}, \ldots, a_{d+1}\right)+\frac{2}{\left[\prod_{j=1}^pa_j\right]}\\&\times\sum_{\mathbf{k}\in(\Z^p\setminus
\{\mathbf{0}\})\times(\Z^{d+1-p}\setminus\{\mathbf{0}\})} \left(\frac{\sum_{j=1}^p\left[\frac{k_j}{a_j}\right]^2}
{\sum_{j=p+1}^{d+1}[a_{j}k_{j}]^2}\right)^{-\frac{p}{4}}K_{\frac{p}{2}}\left(2\pi\sqrt{\left(
\sum_{j=1}^p\left[\frac{k_j}{a_j}\right]^2
\right)\left(\sum_{j=p+1}^{d+1}[a_{j}k_{j}]^2\right)}\right).
\end{split}\end{equation}From this, we obtain the following alternative expressions for the regularized Casimir energy $E_{\text{Cas, reg}}^P(L_1, \ldots, L_d)$ \eqref{eq10_8_1}
  for
massless scalar field with Pb.c.:
\begin{align}\label{eq7_22_3}
&E_{\text{Cas, reg}}^P(L_1, \ldots, L_d) = T\log L_1 + T\log T+\pi T
L_1 Z_d\left(-\frac{1}{2}; \frac{1}{L_2}, \ldots, \frac{1}{L_d},
T\right)\\
&-T\sum_{k_1=1}^{\infty} \sum_{(k_2, \ldots,
k_{d}, l)\in\Z^{d}\setminus\{0\}}k_1^{-1}\exp\left(-2\pi k_1 L_1 \sqrt{\sum_{j=2}^d
\left(\frac{k_j}{L_j}\right)^2 +(lT)^2}\right).\nonumber
\end{align}
\begin{align}\label{eq7_14_8}
&E_{\text{Cas, reg}}^P(L_1,  \ldots, L_d) =-\frac{T}{2}Z_{E, d}'\left(0; T, \frac{1}{L_2}, \ldots,\frac{1}{L_d}\right)-T\log\frac{2\pi}{T}
\\&-\frac{\Gamma\left(\frac{d+1}{2}\right)}{\pi{^\frac{d+1}{2}} L_1^d} \left[\prod_{j=2}^d L_i\right]\zeta_R(d+1)\nonumber
-2L_1^{-\frac{d}{2}}\left[\prod_{j=2}^d L_i\right]\sum_{k_1=1}^{\infty} \sum_{(k_2, \ldots, k_d, l)\in \Z^{d}\setminus\{0\}}
k_1^{\frac{d}{2}}\\&\times\left(\sum_{j=2}^d (k_jL_j)^2 + \left(\frac{l}{T}\right)^2\right)^{-\frac{d}{4}} K_{\frac{d}{2}}\left( \frac{2\pi k_1}{L_1}
\sqrt{\sum_{j=2}^d (k_jL_j)^2 + \left(\frac{l}{T}\right)^2}\right)\nonumber.
\end{align}
\begin{align}\label{eq7_16_4}
&E_{\text{Cas, reg}}^P(L_1, \ldots, L_d) = -\frac{\Gamma\left(\frac{d+1}{2}\right)\zeta_R(d+1)}{\pi^{\frac{d+1}{2}}}\left[\prod_{j=1}^d L_j\right]T^{d+1}
-T\log\frac{2\pi}{T}\\&-\frac{T}{2}Z_d'\left(0; \frac{1}{L_1}, \ldots,\frac{1}{L_d}\right)-2\left[\prod_{j=1}^d L_j\right]T^{\frac{d+2}{2}}
\sum_{l=1}^{\infty}\sum_{\mathbf{k}\in\Z^d\setminus\{0\}}l^{\frac{d}{2}}\nonumber\\&\times \left(\sum_{j=1}^d (k_jL_j)^2\right)^{-\frac{d}{4}}K_{\frac{d}{2}}
\left(2\pi l T\sqrt{\sum_{j=1}^d (k_jL_j)^2}\right)\nonumber
\end{align}
\section{One, two and three dimensional pistons}In this appendix, we present the results we obtained in the previous sections to special cases with $d=1, 2, 3$.

\subsection{$d=1$} As we mention in Section 2, there are no electromagnetic field in dimension $d=1$. For massless scalar field with Pb.c., we find from \eqref{eq7_15_16} that the low temperature expansion of the Casimir force is equal to
\begin{align}\label{eq7_22_8}
F_{\text{Cas}}^P(a) = F_{\text{Cas}}^{P, T=0}(a)-\frac{\pi T^2}{6}+\frac{4\pi}{a^2}\sum_{l=1}^{\infty}\sum_{k=1}^{\infty} k\exp\left(-\frac{2\pi k l}{T a}\right),
\end{align}  where $F_{\text{Cas}}^{P, T=0}(a)$ is the zero temperature Casimir force given by
\begin{align*}
F_{\text{Cas}}^{P, T=0}(a) = -\frac{\pi}{6 a^2}.
\end{align*}For the high temperature expansion of the Casimir force,  eqs. \eqref{eq7_16_3}  gives
\begin{align}\label{eq7_22_9}
F_{\text{Cas}}^P(a) = -\frac{T}{a} -4\pi T^{2}\sum_{k=1}^{\infty}\sum_{l=1}^{\infty}  l e^{-2\pi kl T a}.
\end{align}This formula can also be directly derived from \eqref{eq7_14_1}. \eqref{eq7_22_9} shows that the classical limit of the Casimir force due to massless scalar field with Pb.c.~ is given by $-T/a$.
For massless scalar field with Db.c.~ and Nb.c., we have
$$F_{\text{Cas}}^{D/N}(a) = F_{\text{Cas}}^P(2a).$$It is interesting to note that \eqref{eq7_22_8} and \eqref{eq7_22_9} give us the identity
\begin{align*}
-\frac{\pi}{6} -\frac{\pi z^2}{6} +4\pi \sum_{k=1}^{\infty}\sum_{l=1}^{\infty} e^{-\frac{2\pi kl}{z}}=-z-4\pi z^2\sum_{k=1}^{\infty}\sum_{l=1}^{\infty} l e^{-2\pi kl z}.
\end{align*}

\subsection{$d=2$} In $d=2$ dimension, the Casimir force acting on a piston for massless scalar field with Db.c.~ and electromagnetic field with PMC b.c.~ coincide; and the Casimir force for massless scalar field with Nb.c.~ and electromagnetic field with PEC b.c.~ coincide. Moreover,
\begin{align}\label{eq7_24_1}
F_{\text{Cas}}^{D/PMC}(a; L_2) =\frac{1}{2}\left( F_{\text{Cas}}^P(2a; 2L_2) -F_{\text{Cas}}^P(2a)\right);\\
F_{\text{Cas}}^{N/PEC}(a; L_2) =\frac{1}{2}\left( F_{\text{Cas}}^P(2a; 2L_2) +F_{\text{Cas}}^P(2a)\right).\nonumber
\end{align}Denote $L_2$ by $b$. Using \eqref{eq7_15_16}, we find that the low temperature expansion of the Casimir force acting on the piston for massless scalar  field with Pb.c.~ is
\begin{align*}
F_{\text{Cas}}^P(a; b)=& F_{\text{Cas}}^{P, T=0}(a; b)+\frac{4\pi}{a^3}\sum_{l=1}^{\infty}\sum_{k_1=1}^{\infty}\sum_{k_2=-\infty}^{\infty} k_1^2\left(\sqrt{\left(\frac{k_1}{a}\right)^2+\left(\frac{k_2}{b}\right)^2}\right)^{-1}\\&\times \exp\left(-\frac{2\pi l}{T}\sqrt{\left(\frac{k_1}{a}\right)^2+\left(\frac{k_2}{b}\right)^2}\right)-\frac{\pi T^2}{6}-\frac{4T}{b}\sum_{l=1}^{\infty}\sum_{k=1}^{\infty} kl^{-1}K_1\left(\frac{2\pi kl}{Tb}\right).
\end{align*}When $a$ is small, the zero temperature Casimir force has a representation
\begin{align*}
F_{\text{Cas}}^{P, T=0}(a; b) = -\frac{\zeta_R(3)}{\pi}\frac{b}{a^3} -\frac{\zeta_R(3)}{2\pi} \frac{1}{b^2} +\frac{8\pi b}{a^3}\sum_{k_1=1}^{\infty}\sum_{k_2=1}^{\infty} k_1^2 K_0\left( \frac{2\pi k_1 k_2 b}{a}\right).
\end{align*}When $a$ is large, it has another representation
\begin{align*}
F_{\text{Cas}}^{P, T=0}(a; b) =-\frac{\pi}{6 a^2}-\frac{4}{a b}\sum_{k_1=1}^{\infty}\sum_{k_2=1}^{\infty} \frac{k_2}{k_1}K_1\left(\frac{2\pi k_1 k_2 a}{b}\right) -\frac{8\pi}{b^2}\sum_{k_1=1}^{\infty}\sum_{k_2=1}^{\infty} k_2^2K_0\left(\frac{2\pi k_1 k_2 a}{b}\right).
\end{align*}
 The high temperature expansion of the Casimir force can be computed by \eqref{eq7_16_3} which gives
\begin{align*}
&F_{\text{Cas}}^P(a; b) =-T\left\{ \frac{\pi b}{6 a^2}+\frac{\pi }{6b} -\frac{4\pi b}{a^2}\sum_{k_1=1}^{\infty}\sum_{k_2=1}^{\infty} k_1e^{-\frac{2\pi k_1 k_2 b}{a}}\right\}+4bT^2\sum_{l=1}^{\infty}\sum_{k_1=1}^{\infty}\sum_{k_2=-\infty}^{\infty} l\\&\times\left( (k_1 a)^2+(k_2 b)^2\right)^{-\frac{1}{2}} K_1\left(2\pi lT\sqrt{(k_1a)^2+(k_2 b)^2}\right)-8\pi a^2 bT^3\sum_{l=1}^{\infty}\sum_{k_1=1}^{\infty}\sum_{k_2=-\infty}^{\infty} l^2 k_1^2\\&\times\left( (k_1 a)^2+(k_2 b)^2\right)^{-1} K_2\left(2\pi lT\sqrt{(k_1a)^2+(k_2 b)^2}\right).
\end{align*}The first term gives the classical limit of the Casimir force for massless scalar field with Pb.c.

For electromagnetic field with PEC b.c.~ or PMC b.c., we can use eq. \eqref{eq7_24_1} to obtain the low temperature expansion and high temperature expansion from the corresponding expansions for massless scalar field with Pb.c. For low temperature, we have
\begin{align*}
&F_{\text{Cas}}^{PEC/PMC}(a; b) =F_{\text{Cas}}^{PEC/PMC, T=0}(a; b)+\frac{\pi}{a^3} \sum_{l=1}^{\infty}\sum_{k_1=1}^{\infty}\sum_{k_2\in (\mathbb{N}\cup\{0\})/\mathbb{N}}  k_1^2\\&\times\left(\sqrt{\left(\frac{k_1}{a}\right)^2+\left(\frac{k_2}{b}\right)^2}\right)^{-1} \exp\left(-\frac{\pi l}{T}\sqrt{\left(\frac{k_1}{a}\right)^2+\left(\frac{k_2}{b}\right)^2}\right)-\frac{\pi T^2}{6}\delta \\&-\frac{T}{b}\sum_{l=1}^{\infty}\sum_{k=1}^{\infty} kl^{-1}K_1\left(\frac{\pi kl}{Tb}\right).
\end{align*}Here $\delta^{PEC}=1$ and $\delta^{PMC}=0$, and the zero temperature Casimir force is given by
\begin{align}\label{eq7_24_6}
F_{\text{Cas}}^{PEC/PMC, T=0}(a; b)=&\frac{1}{8} F_{\text{Cas}}^{P, T=0}(a; b)\mp \frac{\pi}{48 a^2}\\
=&-\frac{\zeta_R(3)}{8\pi}\frac{b}{a^3} -\frac{\zeta_R(3)}{16\pi} \frac{1}{b^2} +\frac{\pi b}{a^3}\sum_{k_1=1}^{\infty}\sum_{k_2=1}^{\infty} k_1^2 K_0\left( \frac{2\pi k_1 k_2 b}{a}\right)\mp \frac{\pi}{48 a^2},\nonumber
\end{align}or
\begin{align}\label{eq7_24_7}
F_{\text{Cas}}^{PEC/PMC, T=0}(a; b)=&-\delta\frac{\pi}{24 a^2}-\frac{1}{2a b}\sum_{k_1=1}^{\infty}\sum_{k_2=1}^{\infty} \frac{k_2}{k_1}K_1\left(\frac{2\pi k_1 k_2 a}{b}\right) \\&-\frac{\pi}{b^2}\sum_{k_1=1}^{\infty}\sum_{k_2=1}^{\infty} k_2^2K_0\left(\frac{2\pi k_1 k_2 a}{b}\right).\nonumber
\end{align}
 For electromagnetic field with PEC b.c.~ or equivalently massless scalar field with Nb.c., eqs. \eqref{eq7_24_6} and \eqref{eq7_24_7} agree with the corresponding formulas in \cite{26}. For electromagnetic field with PMC or equivalently massless scalar field with Db.c., they agree with the results of \cite{14, 29}.

For high temperature,
\begin{align*}
&F_{\text{Cas}}^{PEC/PMC}(a; b) =-T\left\{\pm  \frac{1}{4a}+\frac{\pi b}{24 a^2}+\frac{\pi }{24 b} -\frac{\pi b}{a^2}\sum_{k_1=1}^{\infty}\sum_{k_2=1}^{\infty} k_1e^{-\frac{2\pi k_1 k_2 b}{a}}\right\}+2bT^2\\&\times \sum_{l=1}^{\infty}\sum_{k_1=1}^{\infty}\sum_{k_2=-\infty}^{\infty} l\left( (k_1 a)^2+(k_2 b)^2\right)^{-\frac{1}{2}} K_1\left(4\pi lT\sqrt{(k_1a)^2+(k_2 b)^2}\right)-8\pi a^2 bT^3 \sum_{l=1}^{\infty}\sum_{k_1=1}^{\infty}\\&\times\sum_{k_2=-\infty}^{\infty} l^2 k_1^2 \left( (k_1 a)^2+(k_2 b)^2\right)^{-1} K_2\left(4\pi lT\sqrt{(k_1a)^2+(k_2 b)^2}\right) \mp 2\pi T^{2}\sum_{k=1}^{\infty}\sum_{l=1}^{\infty}  l e^{-4\pi kl T a}.
\end{align*}The first term is the classical limit of the Casimir force for electromagnetic field.

In the  $b\rightarrow \infty$ limit, we find that the pressure on the infinite piston $x_1=a$ due to massless scalar field with Pb.c.~ is
\begin{align*}
\mathcal{P}_{\text{Cas}}^P(a) =-\frac{\zeta_R(3)}{\pi a^3}-\frac{T^3}{2\pi}\zeta_R(3)+\frac{8\pi}{a^3}\sum_{k=1}^{\infty}\sum_{l=1}^{\infty}k^2 K_0\left(\frac{2\pi kl}{Ta}\right),
\end{align*}
or
\begin{align*}
\mathcal{P}_{\text{Cas}}^P(a)=-\frac{\pi}{6}\frac{T}{a^2} +\frac{4T^2}{a}\sum_{k=1}^{\infty}\sum_{l=1}^{\infty}lk_1^{-1}K_1(2\pi kl Ta) -8\pi T^3\sum_{k=1}^{\infty}\sum_{l=1}^{\infty} l^2 K_2(2\pi kl Ta).
\end{align*}For massless scalar field with Db.c.~ or Nb.c., $\mathcal{P}_{\text{Cas}}^{D/N}(a) = \mathcal{P}_{\text{Cas}}^P(2a)$.

\subsection{$d=3$}When $d=3$, the Casimir force for the electromagnetic field with PEC b.c.~ and PMC b.c.~ coincide. Moreover,
\begin{align}\label{eq7_24_5}
F_{\text{Cas}}^{D/N}(a; L_2, L_3)=&\frac{1}{4}\left( F_{\text{Cas}}^P(2a; 2L_2, 2L_3) \mp F_{\text{Cas}}(2a; 2L_2) \mp F_{\text{Cas}}^P(2a; 2L_3) + F_{\text{Cas}}^P(2a)\right),\\
F_{\text{Cas}}^{PEC}(a; L_2, L_3)=& 2 F_{\text{Cas}}^D ( a; L_2, L_3)+ F_{\text{Cas}}^D(a; L_2) + F_{\text{Cas}}^D(a; L_3)=\frac{1}{2}\left(F_{\text{Cas}}^P(2a; 2L_2, 2L_3)- F_{\text{Cas}}^P(2a)\right).\nonumber
\end{align}Setting $L_2=b$ and $L_3=c$, we find from \eqref{eq7_15_16} that in the low temperature limit, the Casimir force acting on the piston for massless scalar field with Pb.c.~ is
\begin{align*}
F_{\text{Cas}}^P(a; b, c) =&F_{\text{Cas}}^{P, T=0}(a; b, c)+\frac{4\pi}{a^3}\sum_{l=1}^{\infty}\sum_{k_1=1}^{\infty}\sum_{(k_2, k_3)\in \Z^2} k_1^2\left(\sqrt{\left(\frac{k_1}{a}\right)^2+\left(\frac{k_2}{b}\right)^2+\left(\frac{k_3}{c}\right)^2}\right)^{-1}\\&\times \exp\left(-\frac{2\pi l}{T}\sqrt{\left(\frac{k_1}{a}\right)^2+\left(\frac{k_2}{b}\right)^2+\left(\frac{k_3}{c}\right)^2}\right)-\frac{\pi T^2}{6}\\&-2T\sum_{l=1}^{\infty}\sum_{(k_2, k_3)\in\Z^2\setminus\{0\}} l^{-1}\sqrt{\left(\frac{k_2}{b}\right)^2 +\left(\frac{k_3}{c}\right)^2}K_1\left(\frac{2\pi l}{T}\sqrt{\left(\frac{k_2}{b}\right)^2 +\left(\frac{k_3}{c}\right)^2}\right);
\end{align*}and for massless scalar field with Db.c.~ and Nb.c.~ and electromagnetic field with PEC b.c.~ and PMC b.c.~ is
\begin{align*}
&F_{\text{Cas}}(a; b, c) =F_{\text{Cas}}^{ T=0}(a; b, c)+\frac{\pi}{a^3}\sum_{l=1}^{\infty}\sum_{k_1=1}^{\infty}\sum_{(k_2, k_3)\in (\mathbb{N}\cup\{0\})^2} k_1^2\Lambda(k_2, k_3)\\&\times\left(\sqrt{\left(\frac{k_1}{a}\right)^2+\left(\frac{k_2}{b}\right)^2+\left(\frac{k_3}{c}\right)^2}\right)^{-1} \exp\left(-\frac{\pi l}{T}\sqrt{\left(\frac{k_1}{a}\right)^2+\left(\frac{k_2}{b}\right)^2+\left(\frac{k_3}{c}\right)^2}\right)-\delta \frac{\pi T^2}{6}\\&-T\sum_{l=1}^{\infty}\sum_{(k_2, k_3)\in(\mathbb{N}\cup\{0\})^2\setminus\{0\}} l^{-1}\Lambda(k_2, k_3)\sqrt{\left(\frac{k_2}{b}\right)^2 +\left(\frac{k_3}{c}\right)^2}K_1\left(\frac{\pi l}{T}\sqrt{\left(\frac{k_2}{b}\right)^2 +\left(\frac{k_3}{c}\right)^2}\right).
\end{align*}Here $\Lambda^D(k_2, k_3) = 1$ if and only if both $k_2$ and $k_3$ are nonzero; $\Lambda^{N}(k_2, k_3) =1$ for all $(k_2, k_3)\in (\mathbb{N}\cup\{0\})^2$, $\Lambda^{PEC}(k_2, k_3)=2$ if both $k_2$ and $k_3$ are nonzero, $\Lambda^{PEC}(k_2, k_3)=1$ if exactly one of the $k_2$ or $k_3$ is zero and $\Lambda^{PEC}(k_2, k_3)=0$ if $k_2=k_3=0$. The zero temperature Casimir force for massless scalar field with Pb.c.~ can be expressed as\begin{align*}
&F_{\text{Cas}}^{P, T=0}(a; b, c)=-\frac{\pi^2 bc}{30a^4} -\frac{bc}{2\pi^2}Z_2(2; b, c)+\frac{2\pi bc}{a^3}\sum_{k_1=1}^{\infty}\sum_{(k_2, k_3)\in\Z^2\setminus\{0\}}k_1^2\left((k_2b)^2+(k_3c)^2\right)^{-1/2}\\&\hspace{4cm}\times \exp\left(-\frac{2\pi k_1}{a}\sqrt{(k_2b)^2+(k_3c)^2}\right)\\
=&-\frac{\pi^2 bc}{30a^4} -\frac{\pi^2}{90}\frac{c}{b^3} -\frac{\zeta_R(3)}{2\pi}\frac{1}{c^2}-\frac{4}{b^{\frac{3}{2}}c^{\frac{1}{2}}}\sum_{k_2=1}^{\infty}\sum_{k_3=1}^{\infty}
\left(\frac{k_2}{k_3}\right)^{\frac{3}{2}}K_{\frac{3}{2}}\left(\frac{2\pi k_2 k_3 c}{b}\right)\\&+\frac{2\pi bc}{a^3}\sum_{k_1=1}^{\infty}\sum_{(k_2, k_3)\in\Z^2\setminus\{0\}}k_1^2\left((k_2b)^2+(k_3c)^2\right)^{-1/2}\exp\left(-\frac{2\pi k_1}{a}\sqrt{(k_2b)^2+(k_3c)^2}\right),
\end{align*}or
\begin{align*}
F_{\text{Cas}}^{P, T=0}(a; b, c)=&-\frac{\pi}{6a^2}-\frac{2}{a} \sum_{k_1=1}^{\infty}\sum_{(k_2, k_3)\in \Z^2\setminus\{0\}}k_1^{-1} \sqrt{\left(\frac{k_2}{b}\right)^2+\left(\frac{k_3}{c}\right)^2}K_1\left(2\pi k_1 a \sqrt{\left(\frac{k_2}{b}\right)^2+\left(\frac{k_3}{c}\right)^2}\right)\\
&-4\pi \sum_{k_1=1}^{\infty}\sum_{(k_2, k_3)\in \Z^2\setminus\{0\}} \left(\left(\frac{k_2}{b}\right)^2+\left(\frac{k_3}{c}\right)^2\right)K_0\left(2\pi k_1 a \sqrt{\left(\frac{k_2}{b}\right)^2+\left(\frac{k_3}{c}\right)^2}\right),
\end{align*}  which are suitable for studying the small $a$ and large $a$ behaviors of the Casimir force. For massless scalar field with Db.c.~ and Nb.c., the corresponding formulas are given respectively by
\begin{align}\label{eq7_24_8}
&F_{\text{Cas}}^{D/N, T=0}(a; b, c)
=-\frac{\pi^2 bc}{480 a^4}\pm \frac{b+c}{16\pi a^3}\zeta_R(3) -\frac{\pi^2}{1440}\frac{c}{b^3} \pm \frac{\zeta_R(3)}{32\pi}\left(\frac{1}{b^2}+\frac{1}{c^2}\right)-\frac{\zeta_R(3)}{32\pi}\frac{1}{c^2}-\frac{\pi}{96 a^2}\\&-\frac{1}{4 b^{\frac{3}{2}}c^{\frac{1}{2}}}\sum_{k_2=1}^{\infty}\sum_{k_3=1}^{\infty}
\left(\frac{k_2}{k_3}\right)^{\frac{3}{2}}K_{\frac{3}{2}}\left(\frac{2\pi k_2 k_3 c}{b}\right)\mp \left\{\frac{\pi b}{2a^3}\sum_{k_1=1}^{\infty}\sum_{k_2=1}^{\infty} k_1^2 K_0\left(\frac{2\pi k_1 k_2 b}{a}\right)+ b \longleftrightarrow c\right\}\nonumber\\&+\frac{\pi bc}{8a^3}\sum_{k_1=1}^{\infty}\sum_{(k_2, k_3)\in\Z^2\setminus\{0\}}k_1^2\left((k_2b)^2+(k_3c)^2\right)^{-1/2}\exp\left(-\frac{2\pi k_1}{a}\sqrt{(k_2b)^2+(k_3c)^2}\right)\nonumber
\end{align}and
\begin{align}\label{eq7_24_2}
&F_{\text{Cas}}^{D/N, T=0}(a; b, c)=-\delta\frac{\pi}{24 a^2}-\frac{1}{2a} \sum_{k_1=1}^{\infty}\sum_{(k_2, k_3)\in (\mathbb{N}\cup\{0\})^2\setminus\{0\}}k_1^{-1} \Lambda(k_2, k_3) \sqrt{\left(\frac{k_2}{b}\right)^2+\left(\frac{k_3}{c}\right)^2}\\&\hspace{5cm}\times K_1\left(2\pi k_1 a \sqrt{\left(\frac{k_2}{b}\right)^2+\left(\frac{k_3}{c}\right)^2}\right)\nonumber\\
&-\pi \sum_{k_1=1}^{\infty}\sum_{(k_2, k_3)\in (\mathbb{N}\cup\{0\})^2\setminus\{0\}} \Lambda(k_2, k_3) \left(\left(\frac{k_2}{b}\right)^2+\left(\frac{k_3}{c}\right)^2\right)K_0\left(2\pi k_1 a \sqrt{\left(\frac{k_2}{b}\right)^2+\left(\frac{k_3}{c}\right)^2}\right).\nonumber
\end{align}Here $\delta^D=0$ and $\delta^N=1$. Eq. \eqref{eq7_24_8} agrees with the results of \cite{25, 26}. For electromagnetic field with PEC b.c.~ or PMC b.c., the large $a$ expansion for the Casimir force is still given by \eqref{eq7_24_2} with $\delta^{PEC}=0$; whereas the small $a$ expansion is
\begin{align*}
&F_{\text{Cas}}^{PEC, T=0}(a; b, c)
=-\frac{\pi^2 bc}{240 a^4} -\frac{\pi^2}{720}\frac{c}{b^3}- \frac{\zeta_R(3)}{16\pi} \frac{1}{c^2}+\frac{\pi}{48 a^2}\\&-\frac{1}{2 b^{\frac{3}{2}}c^{\frac{1}{2}}}\sum_{k_2=1}^{\infty}\sum_{k_3=1}^{\infty}
\left(\frac{k_2}{k_3}\right)^{\frac{3}{2}}K_{\frac{3}{2}}\left(\frac{2\pi k_2 k_3 c}{b}\right)\\&+\frac{\pi bc}{4 a^3}\sum_{k_1=1}^{\infty}\sum_{(k_2, k_3)\in\Z^2\setminus\{0\}}k_1^2\left((k_2b)^2+(k_3c)^2\right)^{-1/2}\exp\left(-\frac{2\pi k_1}{a}\sqrt{(k_2b)^2+(k_3c)^2}\right).
\end{align*}This agrees with the results of \cite{15, 16, 29}.

For the high temperature expansion of the Casimir force acting on the piston due to massless scalar field with Pb.c., we obtain from \eqref{eq7_16_3} the formula
\begin{align}\label{eq7_27_1}
&F_{\text{Cas}}^P(a; b, c)= -T\Biggl\{\frac{1}{a} +2\pi \sum_{k_1=1}^{\infty}\sum_{(k_2, k_3)\in\Z^2\setminus\{0\}}\sqrt{\left(\frac{k_2}{b}\right)^2+\left(\frac{k_3}{c}\right)^2}\\&\times \exp\left(-2\pi k_1 a \sqrt{\left(\frac{k_2}{b}\right)^2+\left(\frac{k_3}{c}\right)^2}\right)\Biggr\}+4bc T^{\frac{5}{2}}\sum_{l=1}^{\infty}\sum_{k=1}^{\infty} \sum_{(k_2, k_3)\in \Z^2}l^{\frac{3}{2}}\left((k_1a)^2+(k_2b)^2+(k_3c)^2\right)^{-\frac{3}{4}}\nonumber\\
&\times K_{\frac{3}{2}}\left(2\pi lT \sqrt{(k_1a)^2+(k_2b)^2+(k_3 c)^2}\right) -8\pi a^2 bc T^{\frac{7}{2}}\sum_{l=1}^{\infty}\sum_{k=1}^{\infty} \sum_{(k_2, k_3)\in \Z^2}l^{\frac{5}{2}}k_1^2\nonumber\\
&\times \left((k_1a)^2+(k_2b)^2+(k_3c)^2\right)^{-\frac{5}{4}}K_{\frac{5}{2}}\left(2\pi lT \sqrt{(k_1a)^2+(k_2b)^2+(k_3 c)^2}\right).\nonumber
\end{align}The Bessel functions $K_{3/2}(z)$ and $K_{5/2}(z)$ can be re-expressed by elementary functions using the formulas
\begin{align*}
K_{\frac{3}{2}}(z) =\sqrt{\frac{\pi}{2z}}\left(1+\frac{1}{z} \right), \hspace{1cm}K_{\frac{5}{2}}(z) =\sqrt{\frac{\pi}{2z}}\left(1+\frac{3}{z}+\frac{3}{z^2}\right).
\end{align*}The high  temperature expansion of the Casimir force for massless scalar field with Db.c.~ and Nb.c.~ and electromagnetic field with PEC b.c.~ or PMC b.c.~ can be obtained using \eqref{eq7_24_5}. The first term in \eqref{eq7_27_1} give the classical term of the Casimir force for massless scalar field with Pb.c. It has an alternative expression given by \eqref{eq7_18_2}:
\begin{align*}
&F_{\text{Cas}}^{P, \text{classical}}(a; b, c)=-T\Biggl\{\frac{\zeta_R(3)}{\pi}\frac{bc}{a^3} + \frac{bc}{4\pi}Z_2\left(\frac{3}{2}; b, c\right) \\&-\frac{4\pi bc}{a^3}\sum_{k_1=1}^{\infty}\sum_{(k_2, k_3)\in\Z^2\setminus\{0\}}k_1^2K_0\left(\frac{2\pi k_1}{a}\sqrt{(k_2 b)^2+(k_3 c)^2}\right)\Biggr\}\\
=&-T\Biggl\{\frac{\zeta_R(3)}{\pi}\frac{bc}{a^3} + \frac{\zeta_R(3)}{2\pi}\frac{c}{b^2}+\frac{\pi}{6c}+\frac{4}{b}\sum_{k_2=1}^{\infty}\sum_{k_3=1}^{\infty} \frac{k_2}{k_3}K_1\left(\frac{2\pi k_2 k_3 c}{b}\right)\\&-\frac{4\pi bc}{a^3}\sum_{k_1=1}^{\infty}\sum_{(k_2, k_3)\in\Z^2\setminus\{0\}}k_1^2K_0\left(\frac{2\pi k_1}{a}\sqrt{(k_2 b)^2+(k_3 c)^2}\right)\Biggr\}.
\end{align*}For massless scalar field with Db.c.~ and Nb.c., the classical term of the Casimir force is
\begin{align*}
&F_{\text{Cas}}^{D/N, \text{classical}}(a; b, c)=-T\Biggl\{\frac{\zeta_R(3)}{8\pi}\frac{bc}{a^3} \mp \frac{\pi(b+c)}{48 a^2} +\frac{1}{8a}+ \frac{\zeta_R(3)}{16\pi}\frac{c}{b^2}+\frac{\pi}{48c}\mp \frac{\pi}{48}\left(\frac{1}{b}+\frac{1}{c}\right) \\& \pm \frac{\pi b}{2a^2}\sum_{k_1=1}^{\infty}\sum_{k_2=1}^{\infty} k_1 e^{-\frac{2\pi k_1k_2 b}{a}}\pm \frac{\pi c}{2a^2}\sum_{k_1=1}^{\infty}\sum_{k_3=1}^{\infty} k_1 e^{-\frac{2\pi k_1k_3 c}{a}}+\frac{1}{2b}\sum_{k_2=1}^{\infty}\sum_{k_3=1}^{\infty} \frac{k_2}{k_3}K_1\left(\frac{2\pi k_2 k_3 c}{b}\right)\\ & -\frac{\pi bc}{2 a^3}\sum_{k_1=1}^{\infty}\sum_{(k_2, k_3)\in\Z^2\setminus\{0\}}k_1^2K_0\left(\frac{2\pi k_1}{a}\sqrt{(k_2 b)^2+(k_3 c)^2}\right)\Biggr\}.
\end{align*}For electromagnetic field with PEC b.c.~ or PMC b.c., the classical term is given by
\begin{align*}
&F_{\text{Cas}}^{PEC/PMC, \text{classical}}(a; b, c)=-T\Biggl\{\frac{\zeta_R(3)}{\pi}\frac{bc}{4a^3}-\frac{1}{4a} + \frac{\zeta_R(3)}{8\pi}\frac{c}{b^2}+\frac{\pi}{24c}\\&+\frac{1}{b}\sum_{k_2=1}^{\infty}\sum_{k_3=1}^{\infty} \frac{k_2}{k_3}K_1\left(\frac{2\pi k_2 k_3 c}{b}\right)  -\frac{\pi bc}{a^3}\sum_{k_1=1}^{\infty}\sum_{(k_2, k_3)\in\Z^2\setminus\{0\}}k_1^2K_0\left(\frac{2\pi k_1}{a}\sqrt{(k_2 b)^2+(k_3 c)^2}\right)\Biggr\}.
\end{align*}

In the infinite parallel plates limit, i.e. $b=c \rightarrow \infty$, the Casimir pressure on the plates due to massless scalar field with Pb.c.~ is
\begin{align*}
\mathcal{P}_{\text{Cas}}^P(a) = -\frac{\pi^2}{30 a^4} -\frac{\pi^2T^4}{90} +\frac{4\pi T}{a^3}\sum_{k=1}^{\infty}\sum_{l=1}^{\infty}k^2 l^{-1}\exp\left(-\frac{2\pi lk}{Ta}\right)
\end{align*} or
\begin{align*}
\mathcal{P}_{\text{Cas}}^P(a)=&-\frac{\zeta_R(3) T}{\pi a^3} +4\frac{T^{\frac{5}{2}}}{a^{\frac{3}{2}}}\sum_{k=1}^{\infty}\sum_{l=1}^{\infty} l^{\frac{3}{2}}k^{-\frac{3}{2}}K_{\frac{3}{2}}(2\pi kl Ta)-8\pi\frac{T^{\frac{7}{2}}}{a^{\frac{1}{2}}}\sum_{k=1}^{\infty}\sum_{l=1}^{\infty} l^{\frac{5}{2}}k^{-\frac{1}{2}}K_{\frac{5}{2}}(2\pi kl Ta).
\end{align*}For massless scalar field with Db.c.~ or Nb.c., $\mathcal{P}_{\text{Cas}}^{D/N}(a) =\mathcal{P}_{\text{Cas}}^P(2a)$. For electromagnetic field with PEC b.c.~ or PMC b.c.,  $\mathcal{P}_{\text{Cas}}^{PEC}(a) =2\mathcal{P}_{\text{Cas}}^P(2a)$ or more precisely\begin{align*}
\mathcal{P}_{\text{Cas}}^{PEC}(a) = -\frac{\pi^2}{240 a^4} -\frac{\pi^2T^4}{45} +\frac{\pi T}{a^3}\sum_{k=1}^{\infty}\sum_{l=1}^{\infty}k^2 l^{-1}\exp\left(-\frac{\pi lk}{Ta}\right),
\end{align*} or
\begin{align*}
\mathcal{P}_{\text{Cas}}^P(a)=&-\frac{\zeta_R(3) T}{4\pi a^3} +2\sqrt{2}\frac{T^{\frac{5}{2}}}{a^{\frac{3}{2}}}\sum_{k=1}^{\infty}\sum_{l=1}^{\infty} l^{\frac{3}{2}}k^{-\frac{3}{2}}K_{\frac{3}{2}}(4\pi kl Ta)-8\sqrt{2}\pi\frac{T^{\frac{7}{2}}}{a^{\frac{1}{2}}}\sum_{k=1}^{\infty}\sum_{l=1}^{\infty} l^{\frac{5}{2}}k^{-\frac{1}{2}}K_{\frac{5}{2}}(4\pi kl Ta).
\end{align*}In particular, in the low temperature limit, we find that the the leading order term of the Casimir pressure on a pair of infinitely conducting parallel plates is the zero temperature Casimir pressure given by the well-known result of Casimir \cite{1}, i.e.
\begin{align*}
\mathcal{P}_{\text{Cas}}^{PEC, T=0}(a)= -\frac{\pi^2}{240 a^4}.\end{align*} The leading term of the temperature correction is
$$-\frac{\pi^2T^4}{45},$$
agreeing with the result of \cite{24_3}. In the high temperature limit, the leading term of the Casimir pressure is the classical term
\begin{align*}
\mathcal{P}_{\text{Cas}}^{PEC, \text{classical}}(a)= -\frac{\zeta_R(3) T}{4\pi a^3},
\end{align*}which agrees with the results of \cite{24_1, 24_2, 24_3}.

\end{document}